\documentclass[final]{hubsarxiv}
\graphicspath{{figures/}}

\usepackage{multirow}
\usepackage{longtable}

\newcommand{\startappendix}{%
  \setcounter{section}{0}%
  \renewcommand{\thesection}{\Alph{section}}%
  \setcounter{table}{0}%
  \renewcommand{\thetable}{A\arabic{table}}%
  \renewcommand{\theHtable}{A\arabic{table}}%
  \setcounter{figure}{0}%
  \renewcommand{\thefigure}{A\arabic{figure}}%
  \renewcommand{\theHfigure}{A\arabic{figure}}%
  \setcounter{equation}{0}%
  \renewcommand{\theequation}{A\arabic{equation}}%
  \renewcommand{\theHequation}{A\arabic{equation}}%
}

\title{Large Language Model-Driven Context-Aware Eco-Feedback Generation and Evaluation}
\shorttitle{LLM-driven context-aware eco-feedback}
\shortauthors{Jung and Babon-Ayeng}
\author{Wooyoung Jung\affmark{a,}\corrmark\orcid{0000-0003-2746-4983} \and Prosper Babon-Ayeng\affmark{a}}
\affil{\affmark{a}Department of Civil and Architectural Engineering and Mechanics, University of Arizona, Tucson, AZ, USA}
\corresponding{wooyoung@arizona.edu}
\paperno{2026-003}
\paperdate{September 2026}
\versionnote{v1 (2026-09-02): first release. v2 (2026-09-03): revised. v3 (2026-11-05): re-set in the HUBS Lab paper layout; content unchanged.}
\published{W. Jung, P. Babon-Ayeng, ``Large language model-driven context-aware eco-feedback generation and evaluation,'' \textit{Energy and Buildings} 370 (2026) 118038. \url{https://doi.org/10.1016/j.enbuild.2026.118038}. \copyright{} 2026. This manuscript version is made available under the CC-BY-NC-ND 4.0 license (\url{https://creativecommons.org/licenses/by-nc-nd/4.0/}).}
\license{CC BY-NC-ND 4.0}
\abstract{The objective of this study is to demonstrate the potential of generating
context-aware eco-feedback---eco-feedback that reflects a household's
contextual characteristics alongside its energy use patterns---through a large
language model-integrated framework. Previous studies have introduced
personalized eco-feedback, mostly relying on household energy use patterns;
however, they frequently did not reflect distinct household characteristics,
including their persona or non-negotiable routines, leaving eco-feedback
ineffective and sometimes superficial. To address these limitations, we
introduce a contextual engineering framework that generates eco-feedback using
a self-consistency with chain-of-thought prompt, leveraging household energy
analysis data, utility rate structures, and household characteristic
information. We conducted a rigorous empirical validation and a combinatorial
evaluation analysis to assess this framework systematically. The former tested
the framework's ability to generate accurate and contextually grounded
eco-feedback for three households by comparing its output against reference
interventions independently derived from the same household data. The latter
examined the framework's adaptability across 400 scenarios spanning 50
households, two utility rate structures, and four behavioral personas. Key
findings were the following: our proposed framework generated eco-feedback
that aligned with reference interventions at a mean rate of 92.0\% and
grounded its recommendations in the provided household data with 95.7\%
citation accuracy. It also proved highly adaptive, shifting both the
appliances targeted and the energy-saving strategies recommended in response
to rate structure and household context. Ultimately, this study contributes to
realizing the next level of context-aware interactions between occupants and
buildings which paves the way for higher occupant living quality and
sustainability.}

\begin{document}
\maketitle

\section{Introduction}
\label{sec:intro}

Homes embody multi-dimensionality \cite{r1}, supporting the daily routines of
residents, offering environments that promote comfort, health, and well-being
\cite{r2}, and shifting their role to serve as workplaces when needed
\cite{r3}. Many of these activities inherently consume energy resources,
positioning homes as a major contributor to total end-use energy consumption
\cite{r4} with complex, diverse usage patterns \cite{r5}. Recently, this
complexity has further intensified with the integration of distributed energy
resources like solar panels and the growing adoption of energy-intensive
appliances like electric vehicle (EV) chargers. In responding to this demand
shift from homes, utilities are increasingly employing demand-side management
strategies such as time-of-use (TOU) pricing \cite{r6,r7,r8} or smart
thermostat incentive programs \cite{r9,r10,r11}. This transition underscores
the role of households as active participants in the energy ecosystem, not
merely consumers but also contributors to demand response, grid flexibility,
and energy efficiency improvements.

However, the information available to households remains very limited
\cite{r12}. For example, homes equipped with smart meters---over 82\% in
North America in 2024 \cite{r13}---may access daily and weekly accumulated
usage profiles \cite{r14}, potentially serving as eco-feedback (i.e.,
feedback that provides users with information about their energy usage
\cite{r15}). However, these profiles often lack personalized suggestions
tailored to each household's unique energy usage patterns. Another eco-feedback
mechanism in current practices is monthly energy bills charged to households,
which also lack detailed information about when, where, and how energy was
expended \cite{r16}. Building on the hypothesis that households are not fully
aware of the environmental impact of their behaviors, eco-feedback should
function as a learning or experimental tool to enhance their understanding of
energy usage behaviors and potentially encourage sustainable practices
\cite{r12}. In practice, however, such feedback remains generic and inattentive to
households' preferences, convenience, and non-negotiable
routines---factors critical to its effectiveness \cite{r17,r18,r19}.

Offering feasible and effective eco-feedback for households has been a
persistent challenging task, because it requires adaptability and an
understanding that households differ in their preferences, constraints, and
circumstances when adopting such measures. Previous studies relied primarily on
energy usage patterns presented in charts, graphs, and pre-framed texts in user
interfaces to deliver personalized eco-feedback to users (e.g., \cite{r20,r21}).
However, these mechanisms may remain limited in depth and adaptability.

\begin{figure}[!htbp]
  \centering
  \includegraphics[width=0.65\textwidth]{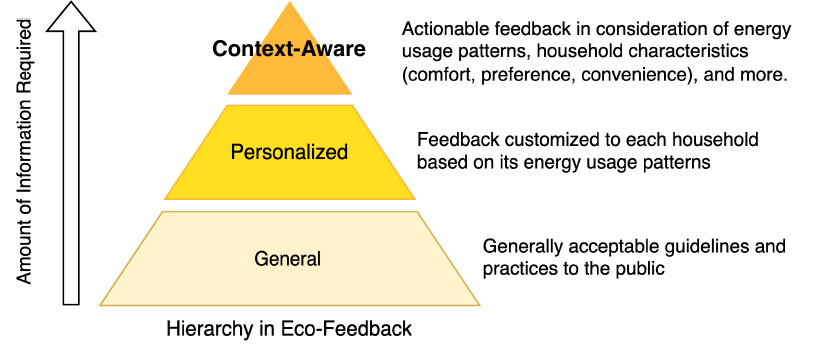}
  \caption{Three Stages of Eco-feedback.}
  \label{fig:hierarchy}
\end{figure}

\begin{figure}[!htbp]
  \centering
  \includegraphics[width=0.65\textwidth]{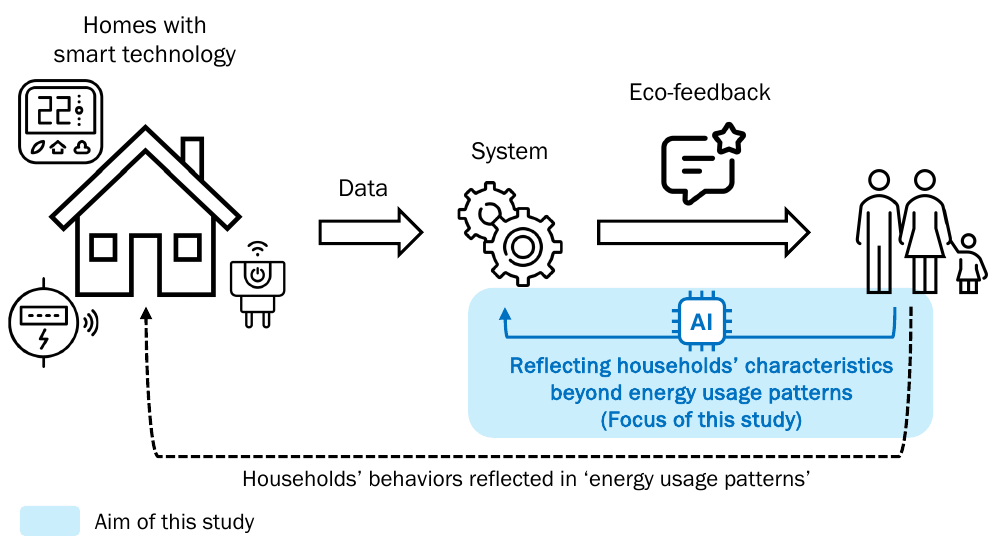}
  \caption{The mechanism of generating context-aware eco-feedback.}
  \label{fig:mechanism}
\end{figure}

The objective of this study is to explore the potential of a large language
model (LLM)-based framework for generating the next stage of eco-feedback,
termed context-aware eco-feedback, which takes both a household's energy
usage patterns and its contextual characteristics into consideration, as
illustrated in Figure~\ref{fig:hierarchy}. LLMs have emerged as cutting-edge artificial intelligence (AI) technology,
recognized for their capability to approximate human-level performance not
only in understanding and generating natural language but also in reasoning
over the information provided to them \cite{r22}. Therefore, we hypothesized that LLMs can effectively
reflect distinct household characteristics, when generating eco-feedback,
extending beyond reliance solely on household energy usage patterns. To address
this hypothesis, this study conducted two complementary analyses using actual
home appliance-level energy use data, utility rate structure data, and
synthesized yet realistic characteristics: (1) an empirical validation
analysis and (2) a combinatorial scenario-based analysis that together revealed both the accuracy
of the approach and its adaptivity to a broader population of households. We
leveraged a context engineering framework, a technique involving optimizing the
information payloads given to LLMs systematically \cite{r23}, as our
methodological foundation to formulate the framework that generates
context-aware eco-feedback. As illustrated in Figure~\ref{fig:mechanism}, this
novel mechanism of generating eco-feedback is expected to facilitate data
integration and information contextualization, enlarging the adaptability in eco-feedback systems in built environments.

% ============================================================
\section{Background}
\label{sec:background}

This section provides background information on (1) the evolution and
limitation of eco-feedback, (2) the integration of LLMs into building energy
management (BEM) tasks and systems, and (3) the foundations of contextual
engineering in LLMs. Together, these areas established the conceptual and
technical basis for this study and highlighted the gaps that motivated our
proposed approach.

\subsection{Eco-Feedback: Evolution and Limitation}
\label{sec:ecofeedback}

Eco-feedback has evolved since the 1970s when studies first monitored
measurable effects \cite{r24}. The early eco-feedback consisted of reporting
the percentage reduction in electricity use on a weekly basis, eventually
serving as a learning tool that allowed users to teach themselves using
experimentation \cite{r12} or a permanent check on the effects of energy
conservation efforts \cite{r25}. Follow-up studies in the 1990s
\cite{r26,r27} discovered that when informative energy bills, incorporating
historic and comparative feedback as well as guidance on which end-uses
consumed the most energy, could influence households' behaviors.

These insights have resulted in interdisciplinary research efforts that
diversified eco-feedback strategies. In designing eco-feedback messages,
studies considered three dimensions: metrics (e.g., energy consumption,
environmental impact), valence (e.g., positive tones), and background
information (e.g., weather, sources for energy, and utility pricing)
\cite{r28}. Specifically, a variety of metrics have been utilized to help
users better understand the extent of their resource consumption. Studies
shared various insights such as users' limited familiarity with energy metrics
(e.g., kilowatt-hour) due to their abstract nature \cite{r29} and their
tendency to interpret actions primarily in terms of costs \cite{r30}.
Score-based indicators, which indirectly reflected resource consumption or
users' actions, were also implemented due to their intuitiveness \cite{r21}.
In terms of valence, Jain et al.\ \cite{r20} observed that users who were
shown positive feedback during their initial login to the user interface were
more likely to continue using their eco-feedback system. Lastly, feedback was
delivered with other background information, such as comparisons with the
user's past performance or with peers \cite{r31}, as well as factors like
weather and utility pricing \cite{r28} to help users understand their resource
consumption.

Another aspect of the evolvement in eco-feedback arose from the adoption of internet-of-things (IoT) technology and high computational
power, which streamlined
the generation of such feedback. Studies have employed smart thermostats, smart
meters, and smart plugs to measure operational and energy consumption data in
real-time with data storage, retrieval, process, and presentation accomplished
instantaneously \cite{r21}. This capability has enabled the design of feedback
strategies that combine timing, frequency, and duration \cite{r28}. Studies
indicated that immediate feedback led to higher adoption rates \cite{r32} and
frequent feedback proved more effective than less frequent feedback \cite{r33}.
However, immediate and frequent feedback could result in diminishing user
engagement or causing feedback fatigue over the long term \cite{r34}.

Despite these advancements, a longstanding limitation of
eco-feedback---one that this study addresses---lies in its limited
adaptivity.
As revealed in \cite{r35}, low adoption rates of eco-feedback often stem from
the systems that fail to account for household characteristics, including their
priorities, non-negotiable routines, or the ones that did not offer feasible
interventions. Consequently, households tend to engage less in sustainable
energy behaviors \cite{r36}. Although personalized eco-feedback partially
mitigated this issue, its reliance on predefined features and energy usage
patterns alone limits its ability to capture the complex and dynamic contexts
that shape household decision-making.

\subsection{Large Language Model-Integrated Building Energy Management}
\label{sec:llm-bem}

The integration of LLMs into BEM has grown rapidly since early 2021 and these research efforts can be clustered
into six thematic areas \cite{r81}: building energy modeling and simulation,
occupant-facing systems and interfaces, HVAC control and thermal comfort, smart 
building and IoT integration, sustainable lifecycle planning, and information 
extraction and automation. Together, these clusters reflect a field in rapid 
transition, moving from early prompt engineering experiments toward domain-adapted 
pipelines incorporating fine-tuning, retrieval-augmented generation (RAG), and 
multi-agent orchestration.

The modeling and simulation cluster illustrates how LLMs can automate
computationally intensive BEM workflows. Studies have applied GPT-based models
to energy load prediction, fault diagnosis, and anomaly detection through
iterative human-model interaction \cite{r37}, and to embed building operational
patterns into automated data mining pipelines via template-based prompt
generation \cite{r38}. Broader explorations have extended LLM application
across EnergyPlus simulation generation, parameter extraction, and model
debugging, demonstrating applicability across the building energy modeling pipeline \cite{r82}.

The occupant-facing systems and interfaces cluster is most directly relevant
to this study. Research here has positioned LLMs as
interfaces between occupants and building systems, including an LLM-based BEM
assistant that combines energy consumption analysis and smart building control
through an API-backed advisory interface \cite{r39}, and a smart home
automation architecture that represents household entities and occupant
preferences as structured textual inputs for LLM-driven management \cite{r40}.
Natural language processing (NLP) and fine-tuned transformer models have also
been applied to classify and
analyze qualitative feedback generated by occupants on indoor environmental
conditions \cite{r83}. Recent studies have further examined how occupants
interact with LLM-integrated systems in real-time settings, investigating user
prompt strategies \cite{r78}, exploring chain-of-thought prompting for
personalized home energy-saving generation \cite{r84}, evaluating
simulation-based assessment methodologies \cite{r79}, and examining how domain
knowledge shapes occupant engagement with AI-driven energy management
\cite{r80}. These applications demonstrate the feasibility of LLM-mediated occupant
engagement with BEM.

Despite this expanding body of work, few studies have targeted the
\textit{generation} of eco-feedback for individual households. Prior
occupant-facing research has concentrated instead on general advisory
settings, single-turn query--response interactions, and system-level control
tasks \cite{r81}. Two requirements therefore remain unaddressed: embedding
household-specific information, such as appliance-level profiles and
structured behavioral personas, into the eco-feedback generation process,
and evaluating output quality across multiple households against defined
behavioral benchmarks. This study addresses both by positioning LLMs as
context-aware generators of household-level eco-feedback, structured around
occupant behavioral profiles and validated through a multi-dimensional
performance evaluation framework.

\subsection{Contextual Engineering in Large Language Models: Methodological
Bridge for Context-Aware Eco-Feedback}
\label{sec:context-eng}

LLMs operate as powerful conditional generators: their outputs depend heavily
on how inputs (collectively referred to as context) are structured,
selected, and framed. Contextual engineering in LLMs refers to the systematic
process of designing an input space by organizing relevant information,
specifying the reasoning scope, and tailoring the prompt in ways that can
optimize model performance \cite{r23}. Rather than treating prompting as an ad
hoc or trial-and-error technique, contextual engineering draws on empirical insights from NLP research demonstrating that models
respond reliably to structured, well-scaffolded context.

Several strands of LLM research provided the foundation for this concept.
Brown et al.\ \cite{r42} demonstrated that the performance of GPT-3 varied
substantially depending on the presence and quality of in-context examples,
establishing prompting as a functional equivalent to task supervision.
Subsequent advances, such as instruction tuning (e.g., \cite{r43}), have
revealed that clearly framed instructions could dramatically increase LLM
robustness and controllability. Work on chain-of-thought prompting
\cite{r44} and self-consistency \cite{r45} revealed that the structure of the
contextual input, such as requiring stepwise reasoning, strongly influenced
model accuracy on tasks involving multi-stage inference. Collectively, these
studies highlighted that LLMs are not just powerful generative models but
context-dependent reasoning systems whose outputs improve when inputs are
carefully engineered.

Contextual engineering also aligns with more recent frameworks that integrate
external data or structured knowledge into prompts. RAG \cite{r46}, for example, showed that providing targeted
factual information enhanced the accuracy of model responses. Similarly,
recent tool-augmented LLM systems \cite{r47} indicated that prompting could
explicitly reference analytical tools or domain-specific computations that
support richer reasoning. These developments revealed how context can function
as an interface between raw domain data and the model's linguistic reasoning
capabilities.

This perspective is particularly relevant to eco-feedback generation in BEM. Eco-feedback is inherently context-sensitive
\cite{r48}: household characteristics, appliance use patterns, variability
constraints, rate structures, and solar generation availability all shape
whether an intervention is appropriate, feasible, or effective. Contextual
engineering allows a model to interpret structured energy data (such as
load profiles, appliance-level metrics, and temporal patterns) alongside
descriptions of household routines or priorities. Accordingly, we expected that when this information is framed coherently
within the prompt, LLMs can generate effective, feasible interventions that
better consider the lived conditions of the household.

Consequently, our hypothesis was that contextual engineering frameworks can
provide the methodological bridge between household contextual data and
adaptive eco-feedback. By curating the information passed to the model and
structuring it in a manner that supports reasoning, this approach can enable
LLMs to serve as context-aware agents capable of producing
household-specific recommendations.

% ============================================================
\section{Methodology}
\label{sec:methodology}

This section presents the proposed conceptual framework to generate
context-aware eco-feedback (Section~\ref{sec:framework}) and details of two
analyses that validated the performance of the framework
(Section~\ref{sec:empirical}) and showed how it can be applied under diverse
household contexts (Section~\ref{sec:combinatorial}). We primarily employed
Python and its libraries to conduct our analyses.

\subsection{Conceptual Framework: Context-Aware Eco-Feedback Generation}
\label{sec:framework}

\begin{figure}[!htbp]
  \centering
  \includegraphics[width=0.70\textwidth]{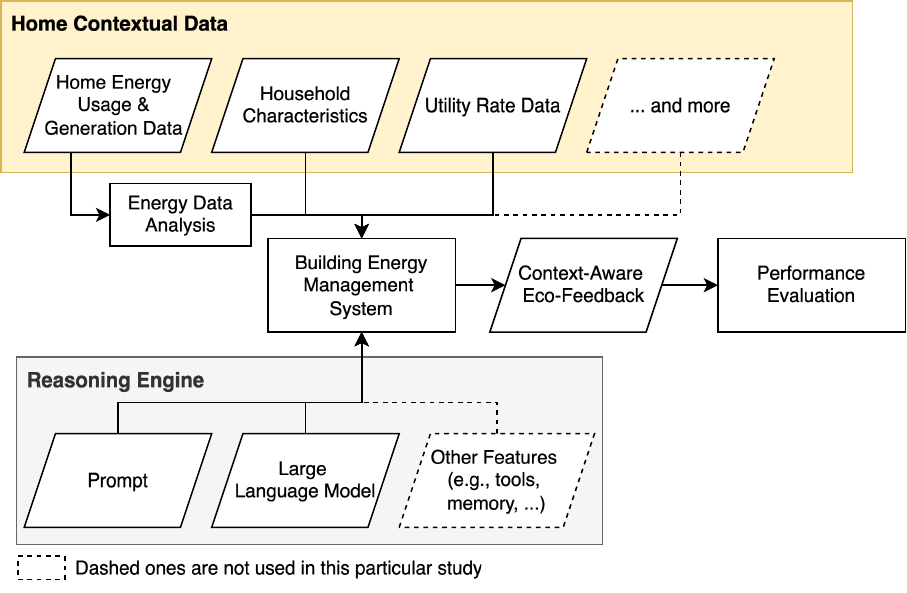}
  \caption{Conceptual framework of generating and validating context-aware
  eco-feedback.}
  \label{fig:framework}
\end{figure}

Figure~\ref{fig:framework} presents the conceptual methodological framework,
presenting how contextual engineering within BEM systems enables the automated
generation of context-aware eco-feedback. It incorporates diverse household
data sources that shape energy-saving opportunities, such as energy usage and
generation data, contextual characteristics (e.g., preferences,
non-negotiable routines), utility rate structures, and others (e.g., building
information such as heating, ventilation, and air-conditioning (HVAC),
lighting systems). These inputs are then transformed into a structured prompt
that the LLM employs to interpret household-specific conditions and generate
context-aware eco-feedback.

The framework integrates four interconnected components. 

\textbf{Household Data}: Diverse data sources can be integrated into this framework when 
they provide meaningful contextual signals. Specifically, high-granularity household 
energy usage data (such as appliance- or circuit-level data) further supports adaptive 
eco-feedback by showing usage patterns, peak contributors, and operational characteristics. 
Utility rate structures supply a critical layer of information by determining if strategies like 
load shifting can be financially beneficial (e.g., under TOU rates) or largely inconsequential 
(e.g., under flat rates). Also, any contextually relevant factors can be included such as 
comfort needs, preferences, routines, constraints, and other behavioral considerations that 
shape the feasibility and acceptance of energy-saving interventions. In addition, building 
information such as window-to-wall ratio, floor plan, HVAC system types can also be 
integrated within this component.

\textbf{Energy Data Analysis}: Building on our previous assessment of prompt engineering to 
analyze household energy usage data \cite{r59}, the energy data analysis component was considered 
within this framework due to a couple of reasons. First, it allows the computationally intensive 
aspects of energy analysis to be handled outside the LLM, relieving the model from tasks that fall outside its primary strengths in NLP
and reasoning. This separation reduces the amount of data (and thus tokens) passed to the
LLM, potentially improving its performance. Second, 
centralizing energy data analysis ensures methodological consistency. Since LLMs may exhibit 
stochastic variation when calculating metrics such as appliance use frequency or variability, 
delegating these computations to a deterministic module helps avoid inconsistencies across runs. 
Finally, if only aggregated household-level energy data are available, this component can 
incorporate load disaggregation methods such as non-intrusive load monitoring \cite{r74} to 
infer appliance-level patterns and support the generation of more granular eco-feedback.

\textbf{Prompt}: Based on user needs or household characteristics, this framework can incorporate 
tailored prompts that guide the LLM toward producing more adoptable and contextually aligned 
eco-feedback. For example, prompts can be adjusted to emphasize cost-saving interventions for 
households prioritizing affordability, highlight comfort-preserving strategies for those with 
limited tolerance for routine disruption, or focus on simple and jargon-free explanations for 
households with lower energy literacy. By structuring prompts to reflect contextual nuances, the 
framework ensures that the LLM's reasoning process remains aligned with household priorities and 
constraints, ultimately yielding eco-feedback that is both feasible and sensitive to the user's 
circumstances.

\textbf{LLM}: This framework can integrate a single or multiple LLM(s) that serve distinct roles 
depending on the design needs of the system. For instance, one LLM may be responsible for 
interpreting contextual inputs and generating eco-feedback, while another may specialize in 
verifying reasoning steps, refining outputs, or ensuring alignment with household constraints. 
Integrating multiple models allows the framework to leverage complementary strengths (such as 
enhanced reasoning, improved reliability, or domain-specific fine-tuning), supporting the generation 
of more accurate, consistent, and context-aware eco-feedback.

A core design principle of the proposed framework is the explicit separation of
deterministic and language-based reasoning. While modern LLMs possess the capacity
to perform arithmetic computations and derive energy metrics from raw data, their
stochastic generation process means that repeated runs on identical inputs do not
guarantee identical analytical outputs (an unacceptable property for energy
metrics that require consistency and reproducibility across evaluation runs).
Delegating all quantitative computations to the deterministic \textit{Energy Data
Analysis} component resolves this: usage frequency, temporal variability, and solar
alignment metrics are computed through fixed, reproducible algorithms, ensuring that
the LLM receives the same structured inputs regardless of when or how many times the
framework is applied. The LLM's role is then restricted to what it is genuinely
suited for: synthesizing these pre-computed metric outputs alongside qualitative
household characteristics (persona priorities, non-negotiable routines, and
utility rate context) into coherent, contextually differentiated eco-feedback.

\subsection{Empirical Validation}
\label{sec:empirical}

\begin{figure}[!htbp]
  \centering
  \includegraphics[width=1.0\textwidth]{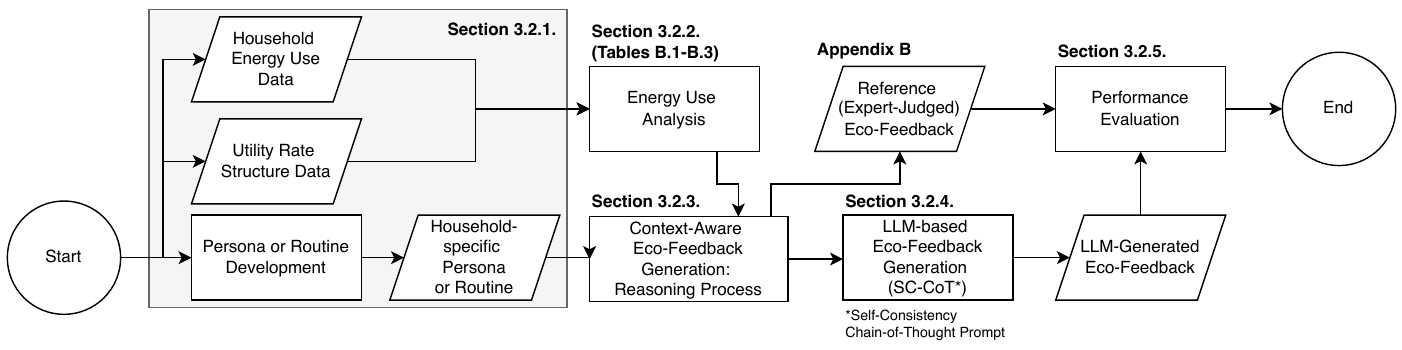}
  \caption{Empirical validation analysis: Methodological overview}
  \label{fig:validation}
\end{figure}

This analysis was designed to validate the framework's ability to generate
context-aware eco-feedback that supports feasible behavior changes that can
effectively result in energy savings. It specifically evaluated if the proposed
framework reliably interpreted household-specific contextual information and
translated it into feasible, high-impact energy-saving recommendations in a
consistent manner. In doing so, as shown in Figure~\ref{fig:validation}, we
compared the context-aware eco-feedback generated using the framework with the
reference interventions for its performance and utilized various performance
evaluation metrics, including data utilization rates, error detection rates, and
consistency. More methodological details are provided below.

\subsubsection{Household data}
\label{sec:emp-data}

This validation analysis utilized three sets of actual appliance-specific home energy usage data collected in
Austin, Texas, USA, combined with two actual utility rates and synthesized
persona depicting unique household preferences or constraints (Table~\ref{tab:homes}).
These three homes were carefully chosen from our 50 household data pool,
purchased from the Pecan Street Inc.\ \cite{r49} (details in
Appendix~\ref{appx:data}), using two criteria: (1) the number of appliances,
which increased the complexity addressed by our framework and (2) the adoption
of solar panels, which diversified the cases included in our analysis. The
inclusion of solar-equipped homes enabled us to assess whether the framework
could recognize energy-saving chances that arise from the presence of on-site
solar generation effectively. We employed two utility rate structures: (1) a
standard rate that applied a uniform cost to total household electricity use
(12.10 cents per kWh) and (2) a TOU rate that introduced financial incentives
for shifting loads away from peak hours (26.10 cents per kWh from 2pm to 8pm
and 9.31 cents during all other hours). TOU rates and on-site solar generation
offer energy-saving opportunities in a similar manner, but they operate through
different mechanisms. Lastly, we considered convenience- or comfort-related
constraints, including thermal comfort, environmental sustainability, and
non-negotiable routines. These constraints were not directly observed but were
incorporated as plausible and evidence-informed assumptions, grounded in
patterns found by previous studies to reflect realistic household
decision-making contexts.

The energy use data consisted of appliance-level power measurements recorded at
15-minute intervals in kilowatts~(kW). Each record contained a timestamp, a
unique household identifier, and per-appliance power consumption readings. For
the solar-equipped home (Home~\#3), household-level photovoltaic generation data
at the same 15-minute resolution were also available. These data covered the
summer months (June through August) in Austin, Texas.

As noted, the raw time-series data were not passed directly to the LLM. Instead, the
deterministic \textit{Energy Data Analysis} component (Section~\ref{sec:framework})
preprocessed all measurements to compute the five appliance-level metrics
described in Section~\ref{sec:metrics}: total energy use and mean power, usage
frequency, usage variability, and solar alignment metrics. These processed metric
outputs, rather than raw interval-level readings, were structured into the
household energy analysis report field of the LLM prompt.
This preprocessing step reduces prompt token volume and ensures that only
analytically relevant, human-interpretable summaries are presented to the model.

\begin{table}[!htbp]
\centering
\caption{Details of three household data}
\label{tab:homes}
\small
\begin{tabular}{p{1.2cm} p{1.7cm} p{2.5cm} p{2.0cm} p{6.5cm}}
\toprule
Home ID & Number of appliances & Solar panel & Utility rate$^{*}$ &
  Persona or non-negotiable routines$^{*}$ \\
\midrule
1 & 7  & Not available & Standard &
  Households in Home~\#1 are sensitive to temperature changes (\cite{r50,r51,r52}). \\
2 & 7  & Not available & TOU &
  Households in Home~\#2 prioritize environmental sustainability over their
  comfort or convenience. In other words, they are passionate about
  energy-saving interventions (\cite{r53}). \\
3 & 14 & Included & Standard &
  All households in Home~\#3 always get together for dinner at 6~pm during
  weekdays to socialize and bond together (\cite{r35,r36,r54}). \\
\bottomrule
\multicolumn{5}{l}{\footnotesize $^{*}$Allocated using the actual utility rate
data and synthesized personas based on literature.}
\end{tabular}
\end{table}

\subsubsection{Energy Use Analysis}
\label{sec:metrics}

To draw reference eco-feedback for these households, we proposed five metrics
to assess each appliance's potential for load curtailment or shiftability.
These metrics were developed after our literature review \cite{r55}.
Their descriptive explanations and mathematical formulations are offered below.
Table~\ref{tab:notation} defines the notation used in the metric equations.

\begin{table}[!htbp]
\centering
\caption{Basic notations and their descriptions}
\label{tab:notation}
\small
\begin{tabular}{l p{11.2cm}}
\toprule
Notation & Description \\
\midrule
$i$            & Appliance type \\
$T$            & Total number of 15-minute intervals per day (96 in our dataset) \\
$t$            & 15-minute interval within the day ($t = 1, 2, \ldots, T$) \\
$h$            & Hour of the day ($h = 0, 1, \ldots, 23$) \\
$d$            & Day index ($d = 1, 2, \ldots, D$) \\
$D$            & Total number of days in the dataset \\
$p$            & Timeframe of interest (e.g., on-peak hours) \\
$H_p$          & Set of hours belonging to timeframe $p$ \\
$|H_p|$        & Number of hours in $H_p$ \\
$N$            & Number of intervals per hour (four in our dataset) \\
$P_{i,d}(t)$   & Power consumption of appliance $i$ at interval $t$ on day $d$ \\
$P_{tot,d}(h)$ & Total household power demand at hour $h$ on day $d$ \\
$P_{PV,d}(h)$  & Solar (PV) power generation at hour $h$ on day $d$ \\
$\tau_i$       & Power threshold separating active and standby modes for appliance $i$ \\
$\tau_{PV}$    & Minimum generation threshold above which solar power is considered available \\
\bottomrule
\end{tabular}
\end{table}

\textbf{1. Total energy and mean power use}: These metrics represented how much
energy (and power) was consumed by an appliance during its usage: the overall
consumption and the mean use. Energy use during standby mode was excluded in
this calculation. Mean power was computed as shown in Equation~(\ref{eq:A8}):
\begin{equation}
  P_i^{avg,p} = \frac{1}{|H_p| \cdot D} \sum_{d=1}^{D} \sum_{h \in H_p}
  P_{i,d}(h)
  \label{eq:A8}
\end{equation}

\noindent where $P_i^{avg,p}$ is the mean power drawn by appliance $i$ during
timeframe $p$, and $P_{i,d}(h)$ is its mean power over hour $h$ on day $d$. The
double sum averages across all hours in $H_p$ and all $D$ days.

\textbf{2. Usage frequency}: This metric captured how frequently an appliance
was used within a timeframe. We considered daily use frequency within both
on- and off-peak hours. An appliance's active status at each 15-minute interval
was determined by the binary indicator function in Equation~(\ref{eq:A1}):
\begin{equation}
  U_{i,d}(t) = \mathbf{1}_{\{P_{i,d}(t) \geq \tau_i\}}
  \label{eq:A1}
\end{equation}

\noindent where $U_{i,d}(t)$ is a binary indicator equal to one when appliance
$i$ is active at interval $t$ on day $d$ and zero otherwise, and
$\mathbf{1}_{\{\cdot\}}$ is the indicator function. The thresholds $\tau_i$ were
identified heuristically to separate active and standby modes \cite{r75,r76}.
The hourly use frequency ($F_{i,d}(h)$) was the sum of binary indicator values
for all 15-minute intervals within an hour (Equation~(\ref{eq:A2})):
\begin{equation}
  F_{i,d}(h) = \sum_{t \in h} U_{i,d}(t)
  \label{eq:A2}
\end{equation}

\noindent where $F_{i,d}(h)$ is the use frequency of appliance $i$ during hour
$h$ on day $d$, and $t \in h$ denotes the four 15-minute intervals falling
within that hour. The normalized average hourly use frequency over $D$ days
($\bar{F}_i^{norm}(h)$) is given by Equation~(\ref{eq:A3}):
\begin{equation}
  \bar{F}_i^{norm}(h) = \frac{1}{D} \sum_{d=1}^{D} \frac{F_{i,d}(h)}{N}
  \label{eq:A3}
\end{equation}

\noindent where $\bar{F}_i^{norm}(h)$ is the average hourly use frequency of
appliance $i$ at hour $h$, normalized to the interval $[0,1]$ by $N$. The
averaged use frequency within a specific timeframe $p$ (e.g., on-peak hours
2:00--8:00~pm) enhanced behavioral interpretability (Equation~(\ref{eq:A4})):
\begin{equation}
  \bar{F}_i^p = \frac{1}{|H_p|} \sum_{h \in H_p} \bar{F}_i^{norm}(h)
  \label{eq:A4}
\end{equation}

\noindent where $\bar{F}_i^p$ is the average normalized use frequency of
appliance $i$ across the hours of timeframe $p$.

\textbf{3. Usage variability}: This metric revealed whether an appliance was
consistently employed at regular times. This use variability metric was motivated
by the consistency metric introduced in \cite{r55}. Unlike their interpretation,
we interpreted households' consistent behaviors as their routines or habits, which
indicated limited variability for change. The higher the use variability, the
more likely households could modify their behavior; appliances combining
significant energy demand with higher variability presented stronger opportunities
for behavioral change (especially load-shifting), whereas consistently used
appliances reflect ingrained habits that are harder to shift. The appliance use
variability ($CV_i$) was calculated through
Equations~(\ref{eq:A5})--(\ref{eq:A7}):
\begin{equation}
  \bar{F}_i = \frac{1}{24} \sum_{h=0}^{23} \bar{F}_i^{norm}(h)
  \label{eq:A5}
\end{equation}
\begin{equation}
  \sigma_i = \sqrt{\frac{1}{24} \sum_{h=0}^{23}
  \left(\bar{F}_i^{norm}(h) - \bar{F}_i\right)^2}
  \label{eq:A6}
\end{equation}
\begin{equation}
  CV_i = \frac{\sigma_i}{\bar{F}_i}
  \label{eq:A7}
\end{equation}

\noindent where $\bar{F}_i$ is the overall mean hourly use frequency of
appliance $i$ across the 24 hours of the day, $\sigma_i$ is the corresponding
standard deviation, and $CV_i$ is the resulting coefficient of variation, with
larger values indicating more variable---and therefore more readily
shiftable---appliance use.

\textbf{4. Solar power alignment}: This metric assessed the extent to which
appliance use aligned with periods of solar power availability, providing an
indication of synchronization between appliance operation and solar production
(Equation~(\ref{eq:A9})):
\begin{equation}
  SPA_i =
  \frac{\displaystyle\sum_{d=1}^{D} \sum_{t=1}^{T}
    \mathbf{1}_{\{P_{i,d}(t)>\tau_i\}} \mathbf{1}_{\{P_{PV,d}(t)>\tau_{PV}\}}}
  {\displaystyle\sum_{d=1}^{D} \sum_{t=1}^{T}
    \mathbf{1}_{\{P_{i,d}(t)>\tau_i\}}}
  \label{eq:A9}
\end{equation}

\noindent where $SPA_i$ is the solar power alignment of appliance $i$. The
numerator counts the intervals in which appliance $i$ is active while solar
generation exceeds $\tau_{PV}$, and the denominator counts all intervals in
which the appliance is active, so that $SPA_i \in [0,1]$.

\textbf{5. Solar power coverage}: This metric quantified the proportion of an
appliance's total energy demand that can be supplied by solar power, reflecting
the U.S.\ convention of net metering, where household-level PV generation was
compared against total aggregated demand at the meter rather than tracked at the
appliance level \cite{r56}. Accordingly, we assumed that each appliance took up
a proportional share of the available solar generation relative to its
instantaneous demand when multiple appliances are operating simultaneously
(Equation~(\ref{eq:A10})):
\begin{equation}
  SPC_i =
  \frac{\displaystyle\sum_{d=1}^{D} \sum_{h \in H_p}
    \left(\frac{P_{i,d}(h)}{P_{tot,d}(h)} \cdot P_{PV,d}(h)\right)}
  {\displaystyle\sum_{d=1}^{D} \sum_{h \in H_p} P_{i,d}(h)}
  \label{eq:A10}
\end{equation}

\noindent where $SPC_i$ is the solar power coverage of appliance $i$. The ratio
$P_{i,d}(h)/P_{tot,d}(h)$ apportions the available solar generation among
simultaneously operating appliances in proportion to their instantaneous demand.

\subsubsection{Context-Aware Eco-feedback Generation: Reasoning Process}
\label{sec:reasoning}

Based on these metrics and other household contextual information, the best
energy-saving interventions through behavioral change were identified within
eco-feedback using the following reasoning steps:

\begin{enumerate}
  \item \textbf{Home Type Assessment}: The utility rate structure and the
        availability of on-site solar energy were considered first to determine
        the home type.
  \item \textbf{Appliance Inventory and Energy Assessment}:
    \begin{itemize}
      \item \textit{Total energy and mean power use}: We assessed the total
            consumption levels to identify appliances that contributed most to
            household demand, therefore highlighting high-use areas where
            behavioral changes could yield meaningful energy savings
            (e.g., $\geq 0.5$\% of the total energy usage). Also, appliances
            with high mean power use were identified since their changes can be
            impactful.
      \item \textit{Usage frequency}: Frequent usage often reflected household
            demand and necessary routines, whereas low frequency with high mean
            energy use might provide strong opportunities for targeted
            interventions, as even small behavioral changes in such cases can
            yield meaningful energy savings.
      \item \textit{Usage variability}: The higher the use variability, the more
            likely households could modify their behavior. In other words,
            appliances that combined significant energy demand with higher
            variability presented stronger opportunities for behavioral change
            (especially load-shifting), whereas consistently used appliances
            reflect ingrained habits that are harder to shift.
      \item \textit{Solar power}: When applicable, two solar power alignment
            metrics (schedule and coverage) were taken into consideration, to
            evaluate whether appliances were used when solar power was available
            or how much energy was derived from solar power.
    \end{itemize}
  \item \textbf{Load Curtailment and Shifting Evaluation}: Building on the
        prior steps, energy-saving strategies through load curtailment or load
        shifting are evaluated systematically. For instance, for households
        subscribing TOU rates, appliances' usage frequency and variability are
        simultaneously considered to assess their load shiftability. Across all
        home types, load curtailment interventions are considered by reducing
        usage frequency, improving efficiency, or substituting with lower-energy
        alternatives.
  \item \textbf{Behavioral Intervention Design with Context}: This step is to
        create effective and feasible energy-saving strategies while considering
        household characteristics, such as their preferences, routines, and
        constraints.
  \item \textbf{Eco-Feedback Finalization}: After considering all previous
        steps, the final energy-saving strategies are included in the
        household's eco-feedback.
\end{enumerate}

These reasoning steps were discussed with an external expert for third-party
adjudication, and their validity was confirmed through a series of discussions.
Consequently, we could identify the effective energy-saving behavioral changes
from the selected households (the detailed analysis results are offered in
Appendix~\ref{appx:reference}). These strategies were treated as reference
interventions for our validation analysis.

\subsubsection{LLM-Incorporated Eco-Feedback Generation}
\label{sec:llm-gen}

\begin{figure}[!htbp]
  \centering
  \includegraphics[width=0.57\textwidth]{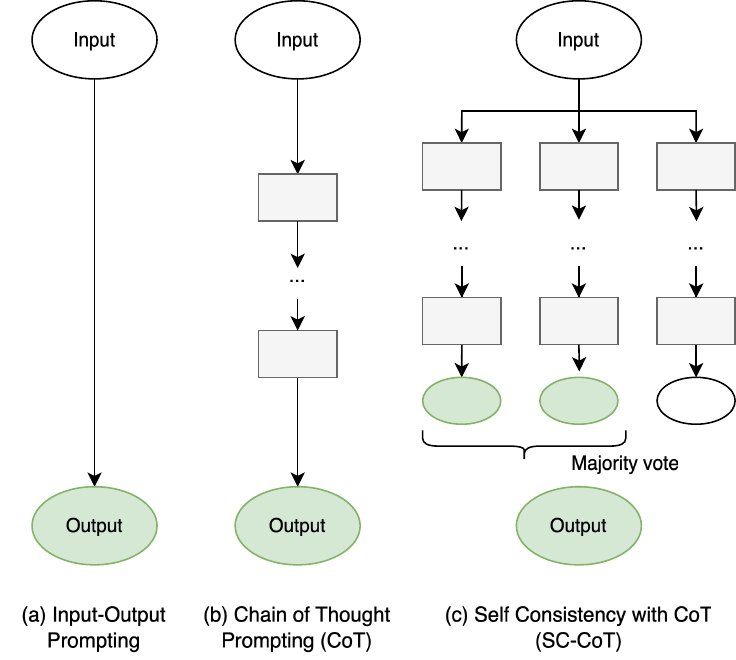}
  \caption{Schematic illustration of different prompting approaches \cite{r57}.}
  \label{fig:sccot}
\end{figure}

We generated context-aware eco-feedback using the framework, illustrated in
Figure~\ref{fig:framework}. The household data presented in
Section~\ref{sec:emp-data} and the eco-feedback generation approach in
Section~\ref{sec:reasoning} were incorporated. Specifically, the five metrics
were implemented into the Energy Data Analysis component and the reasoning
steps were integrated through a prompt. Specifically, we employed
self-consistency with chain-of-thought (SC-CoT) prompt \cite{r45}. SC-CoT
prompt utilizes chain-of-thought (CoT) prompting \cite{r44}, a technique that
uses a series of intermediate reasoning steps to reach a final answer,
multiple times and returns the most frequent outcomes
(Figure~\ref{fig:sccot}).

\begin{figure}[!htbp]
  \centering
  \includegraphics[width=0.95\textwidth]{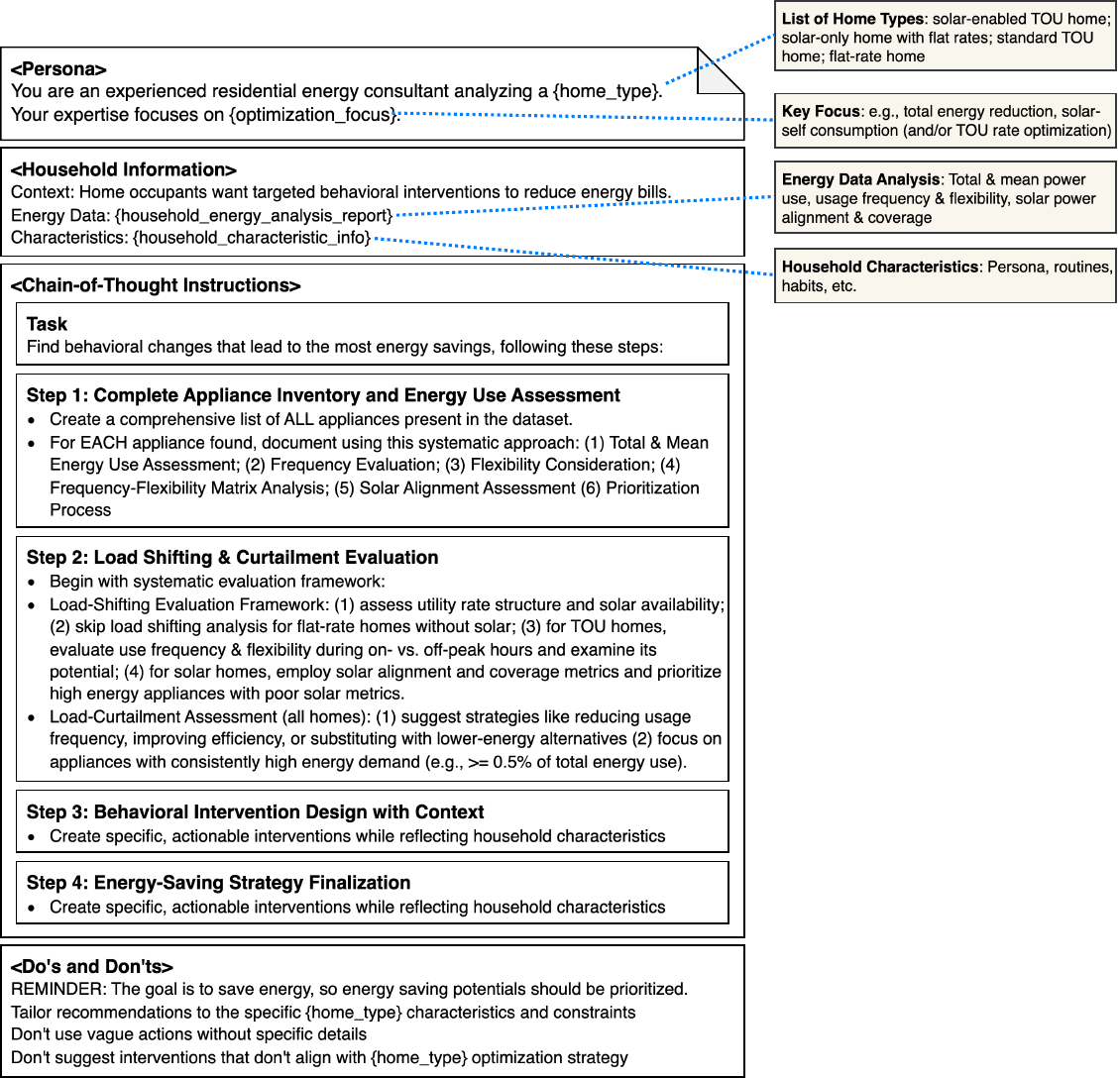}
  \caption{The SC-CoT prompt employed in this study.}
  \label{fig:prompt}
\end{figure}

This technique employs the intuition that complex reasoning tasks typically
admit multiple reasoning paths that reach a correct answer \cite{r58}.
Ultimately, SC-CoT prompt enhances reliability by consolidating reasoning paths
and favoring the solution that consistently emerges across multiple attempts.
Therefore, SC-CoT prompt was an appropriate choice for this validation analysis,
since it allowed multiple reasoning steps that could holistically incorporate
diverse contextual factors and various home energy analysis metrics, while also
improving consistency across the generated solutions. To operationalize this
technique, we retained the final set of AI-generated strategies that appeared
at least three times across five independent runs, ensuring that only the most
stable and recurrent were carried forward. The overview of our SC-CoT prompt
is provided in Figure~\ref{fig:prompt}. This prompt reflected the reasoning
steps employed to establish reference eco-feedback (described in
Section~\ref{sec:reasoning}), but restructured them into a more explicit and
systematic format that enabled the LLM to generate coherent, consistent, and
context-aware eco-feedback. Lastly, we utilized the OpenAI GPT~4o model as the
engine for generating context-aware eco-feedback considering its reasoning
capability and reasonable pricing.

\subsubsection{Performance Evaluation}
\label{sec:perf-eval}

Utilizing this framework, we generated context-aware eco-feedback 50 times
for each household. Then, the performance of the proposed framework was examined
using the following metric categories (detailed corresponding equations are
provided after their descriptive explanations):

\begin{itemize}
  \item \textbf{Accuracy}: This category assesses the correctness of
        LLM-generated eco-feedback from two complementary perspectives:
    \begin{itemize}
      \item \textit{Appliance validation rate} assesses the proportion of
            interventions that target appliances actually present in the
            household's data (Equation~(\ref{eq:avr})).
      \item \textit{Reference alignment rate} measures the proportion of
            interventions that match reference interventions within the
            generated eco-feedback (Equation~(\ref{eq:rar})).
    \end{itemize}
  \item \textbf{Data citation}: This category evaluates how effectively
        LLM-generated interventions reference the provided household data in
        their responses across five data categories: appliance energy use,
        appliance usage patterns, utility rate structure, household context, and
        solar analysis. In other words, this category quantifies the degree to
        which LLM-generated interventions were anchored in actual household data
        rather than generic advice or fabricated information.
    \begin{itemize}
      \item \textit{Data citation frequency} measures how often each data
            category is referenced across all interventions, indicating the
            model's tendency to incorporate specific data types into its
            recommendation (Equation~(\ref{eq:dcf})).
      \item \textit{Data citation accuracy} measures the correctness of
            references when a data category is cited (Equation~(\ref{eq:dca})).
    \end{itemize}
  \item \textbf{Token efficiency}: This category evaluates how effectively the
        LLM utilizes tokens (both input and output) to generate feasible and
        effective household-specific energy-saving recommendations. In
        API-based LLM deployments, token usage directly correlates with
        computational cost and response latency, making efficient token
        utilization a practical consideration for real-world applications.
        Beyond cost implications, token efficiency serves as a proxy for both
        context engineering effort (input) and response quality (output).
        Accordingly, three complementary metrics are considered:
    \begin{itemize}
      \item \textit{Input tokens} measure prompt length, reflecting the
            context engineering effort required to provide the LLM with
            household-specific data including appliance energy consumption,
            usage patterns, utility rate structures, and household
            characteristics.
      \item \textit{Mean output tokens} establish the baseline response
            length, measuring how many tokens the model typically generates in
            its eco-feedback.
      \item \textit{Average output tokens per reference-aligned intervention}
            indicate how many tokens are consumed to generate each
            intervention that aligns with reference interventions
            (Equation~(\ref{eq:tv})).
    \end{itemize}
\end{itemize}

The following are the equations of the metrics stated above.
\begin{equation}
  V = \frac{A_{\text{aligned}}}{A_{\text{total}}}
  \label{eq:avr}
\end{equation}

\noindent where $V$ is the appliance validation rate, $A_{\text{total}}$ is the
number of appliances targeted by the generated interventions, and
$A_{\text{aligned}}$ is the number of those appliances present in the
household's metered data.
\begin{equation}
  R = \frac{I_{\text{aligned}}}{I_{\text{total}}}
  \label{eq:rar}
\end{equation}

\noindent where $R$ is the reference alignment rate, $I_{\text{total}}$ is the
total number of interventions in the generated eco-feedback, and
$I_{\text{aligned}}$ is the number of those interventions matching a reference
intervention. For each appliance, at most one generated intervention could be
counted as aligned, preventing over-counting while ensuring strategic validity.
\begin{equation}
  D_f(c) = \frac{I_{\text{cited}}(c)}{I_{\text{total}}}
  \label{eq:dcf}
\end{equation}

\noindent where $D_f(c)$ is the data citation frequency for category $c$, and
$I_{\text{cited}}(c)$ is the number of interventions referencing data category
$c$, regardless of whether the reference is correct. The categories $c$ comprise
appliance energy use, appliance usage patterns, utility rate structure,
household context, and solar analysis.
\begin{equation}
  D_a(c) = \frac{I_{\text{grounded}}(c)}{I_{\text{cited}}(c)}
  \label{eq:dca}
\end{equation}

\noindent where $D_a(c)$ is the data citation accuracy for category $c$, and
$I_{\text{grounded}}(c)$ is the number of interventions referencing data
category $c$ correctly. Accuracy is conditional on citation: an intervention
that did not reference category $c$ contributes to neither term.

\noindent Input tokens ($T_i$) and mean output tokens ($\bar{T}_o$): Token
counts were obtained directly from the OpenAI API response, which reported the
exact number of input tokens and output tokens, as calculated by the model's
tokenizer.
\begin{equation}
  T_v = \frac{T_o}{I_{\text{aligned}}}
  \label{eq:tv}
\end{equation}

\noindent where $T_v$ is the average number of output tokens per
reference-aligned intervention, $T_o$ is the total number of output tokens
generated across the 50 runs, and $I_{\text{aligned}}$ is the total number of
reference-aligned interventions across those same runs.

We utilized a two-step analysis approach to facilitate this evaluation process.
The first step utilized the LLM, the OpenAI GPT-4o-mini model, to assess the
performance metrics and generate accompanying reasoning traces, allowing us to
capture why specific outcomes were derived. This model was used in our previous
work \cite{r59} and revealed its capability in the evaluation process as a
supportive role. After that, we manually validated the LLM-based performance
evaluations to ensure their credibility. When incorrect interpretations were
provided by the LLM, we manually corrected the scores.

\subsection{Combinatorial Analysis}
\label{sec:combinatorial}

\begin{figure}[!htbp]
  \centering
  \includegraphics[width=0.92\textwidth]{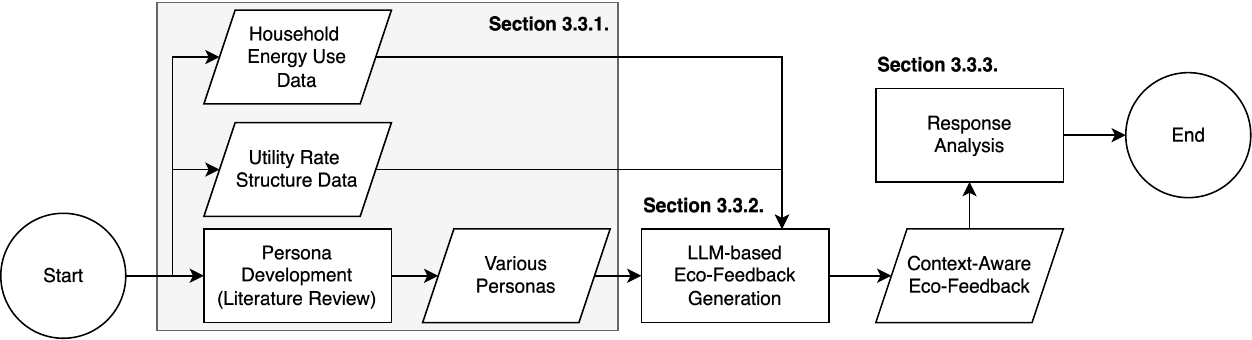}
  \caption{Combinatorial analysis: workflow}
  \label{fig:scalability}
\end{figure}

The objective of this analysis was to assess the proposed framework's
adaptability by applying it to broader household cases.
Figure~\ref{fig:scalability} illustrates how this combinatorial
scenario-based analysis was conducted.

\subsubsection{Household Data}
\label{sec:comb-data}

This analysis included 50 actual household energy usage data (more details are
provided in Appendix~\ref{appx:data}). To generate diverse cases, we
considered two actual utility rate structures: standard flat and TOU rates that
were utilized in the empirical validation analysis. Household characteristics can be broadly diverse, so we developed four personas.
A \textit{persona} is a synthesized behavioral profile constructed from prior
behavioral and energy research that characterizes a household's motivational
priorities, lifestyle constraints, and decision-making tendencies with respect
to energy use. Each persona captures distinct behavioral drivers---such as cost
sensitivity, comfort dependence, technology orientation, or fixed daily
routines---that collectively determine what categories of energy-saving
interventions a household would find feasible and acceptable. Unlike demographic
profiles, personas are designed to represent behavioral archetypes that transcend
any single household, enabling systematic assessment of how eco-feedback adapts
to the diversity of household decision-making contexts. Four personas were
developed based on a structured review of prior residential energy behavior
research, as presented in Table~\ref{tab:personas}.

\begin{table}[!htbp]
\centering
\caption{Synthesized personas developed for the combinatorial analysis}
\label{tab:personas}
\small
\begin{tabular}{p{2.2cm} p{9.5cm} p{1.2cm}}
\toprule
Persona & Description & Ref. \\
\midrule
Cost-minimizers &
  Prioritizing reducing monthly utility expenses and is highly responsive to
  pricing signals. & \cite{r60,r61} \\
Comfort-maximizer &
  Placing strong emphasis on maintaining thermal and lifestyle comfort. This
  persona is less willing to adjust HVAC setpoints. & \cite{r62,r63,r64} \\
Tech-adopter &
  Eager to use automation, combination of smart home integrations, and device
  settings to save energy. & \cite{r65,r66} \\
Routine-constrainer &
  Having strict routine constraints. Interventions must respect immovable tasks
  (meal prep, bath time, nursing schedules) and focus on areas with less time
  sensitivity. & \cite{r67,r68,r69} \\
\bottomrule
\end{tabular}
\end{table}

\subsubsection{Eco-Feedback Generation}
\label{sec:comb-gen}

These three factors resulted in 400 scenarios (50 households, two utility rate
structures, and four personas). Each scenario represented a unique combination
of actual appliance-level energy usage patterns, utility rate conditions, solar
power availability, and contextual characteristics. Using the framework,
validated in Section~\ref{sec:empirical}, we generated three responses for each case (a total of 1200 responses) for comprehensive analysis.

\subsubsection{Response Analysis}
\label{sec:comb-analysis}

The collected responses were examined using the following metrics:

\begin{itemize}
  \item \textbf{Recommended appliance}: We identified the appliances
        recommended across all households and their frequencies.
  \item \textbf{Intervention type}: We classified intervention types using
        three mutually exclusive categories. \textit{Shifting} referred to
        interventions that moved appliance operation to a different time
        without changing the quantity of energy used. \textit{Curtailment}
        referred to interventions that reduced the amount or duration of
        appliance use. \textit{Efficiency} referred to interventions that
        aimed to improve appliance efficiency through settings or upgrades
        rather than reduction or scheduling changes. Each intervention was
        assigned to exactly one category.
  \item \textbf{Persona fidelity score}: This metric evaluated how well
        LLM-generated interventions reflected the behavioral characteristics
        defined by the assigned persona. Specifically, each response was
        assessed using persona-specific criteria in Table~\ref{tab:criteria}
        and quantified based on its alignment. We employed a two-step evaluation
        approach. The first step was an LLM-as-judge method, where the OpenAI
        GPT~4o-mini model rated each response on the four criteria via a
        three-point scale: 0.0 (not present), 0.5 (partially present), and
        1.0 (strongly present). After that, we manually validated all points.
        The final score was calculated as the arithmetic mean of the four
        criterion ratings, scaled to a 0--100 range. Equal weighting was
        applied across all criteria. This approach enabled systematic
        quantification of how faithfully LLM outputs embodied the priorities
        and preferences characteristic of each persona type.
\end{itemize}

\begin{longtable}{p{2.2cm} p{13.4cm}}
\caption{Four evaluation criteria for persona fidelity score}
\label{tab:criteria}\\
\toprule
Persona & Evaluation criteria (with behavioral rationale) \\
\midrule
\endfirsthead
\multicolumn{2}{c}{\textit{(Table~\ref{tab:criteria} continued)}}\\
\toprule
Persona & Evaluation criteria (with behavioral rationale) \\
\midrule
\endhead
\bottomrule
\endlastfoot
Cost-minimizer &
  1.~\textit{Cost sacrifice acceptance}: whether the response stated that comfort
  or convenience tradeoffs were acceptable or worthwhile for cost savings?
  Captures a defining characteristic of price-sensitive households who accept
  minor sacrifices in exchange for meaningful bill reductions \cite{r60,r61}.\newline
  2.~\textit{Equipment upgrade skepticism}: whether the response avoids or
  deprioritizes equipment replacements, favoring low-cost behavioral changes?
  Reflects cost-minimizers' preference for no-capital-investment strategies
  over hardware upgrades \cite{r60}.\newline
  3.~\textit{Aggressive reduction targets}: whether the response recommended
  usage reduction of 20\% or more for high-consuming appliances? Reflects
  cost-minimizers' preference for high-impact curtailment to maximize
  savings \cite{r61}.\newline
  4.~\textit{Comparative cost ranking}: whether the response ranked or compared
  interventions by cost-effectiveness or economic values? Reflects
  cost-minimizers' primary decision criterion of financial impact \cite{r60,r61}. \\[4pt]
Comfort-maximizer &
  1.~\textit{Comfort preservation}: whether the response explicitly prioritized
  maintaining comfort? Captures the core constraint of comfort-maximizers, for
  whom thermal and lifestyle quality are non-negotiable priorities \cite{r62,r63}.\newline
  2.~\textit{Gentle HVAC changes}: whether the response suggested moderate
  ($\leq$1\textdegree C) adjustments or avoided HVAC changes? Reflects
  comfort-maximizers' sensitivity to temperature changes and resistance to
  significant setpoint adjustments \cite{r62,r63}.\newline
  3.~\textit{No sacrifice language}: whether the response avoided asking for
  behavioral sacrifices? Captures comfort-maximizers' unwillingness to accept
  reduced quality of life as a condition of energy savings \cite{r63,r64}.\newline
  4.~\textit{Efficiency focus}: whether the response suggested efficiency
  improvements over curtailment? Reflects comfort-maximizers' preference for
  maintaining usage levels while improving appliance performance \cite{r64}. \\[4pt]
Tech-adopter &
  1.~\textit{Automation suggestions}: whether the response recommended automated
  or smart solutions? Captures tech-adopters' enthusiasm for passive,
  device-driven energy management \cite{r65,r66}.\newline
  2.~\textit{Scheduling emphasis}: whether the response suggested programmable
  schedules or timers? Reflects tech-adopters' reliance on device-level
  automation to optimize energy use without manual intervention \cite{r65}.\newline
  3.~\textit{Smart device integration}: whether the response mentioned smart
  thermostats, apps, or connected devices? Captures tech-adopters' expectation
  that recommendations leverage accessible smart home technology \cite{r65,r66}.\newline
  4.~\textit{Technology-enabled saving}: whether the response stated savings
  achieved through technology rather than manual effort? Reflects tech-adopters'
  preference for passive savings mechanisms \cite{r66}. \\[4pt]
Routine-constrainer &
  1.~\textit{Name constraint examples}: whether the response named specific
  constraints that might exist (e.g., meal prep, bedtime routine, school
  schedule)? Captures routine-constrainers' need for context-sensitive guidance
  that acknowledges schedule specificity \cite{r67,r68}.\newline
  2.~\textit{Flexible window specificity}: whether the response provided specific
  flexible time windows (mid-morning, mid-afternoon) distinct from constrained
  times (breakfast, dinner, bedtime)? Reflects routine-constrainers' need for
  temporally precise recommendations \cite{r68,r69}.\newline
  3.~\textit{Schedule inquiry proactivity}: whether the response acknowledged
  specific household routines, schedules, or timing constraints? Captures
  routine-constrainers' expectation that eco-feedback respect their established
  daily structure \cite{r67,r69}.\newline
  4.~\textit{Time-constrained appliance avoidance}: whether the response avoided
  recommending changes to appliances with inherently constrained timing (cooktop,
  oven)? Reflects routine-constrainers' non-negotiable usage patterns around
  meal preparation and daily rituals \cite{r68,r69}. \\
\end{longtable}

With these metrics, we conducted descriptive and inferential analysis to
capture (1) how intervention characteristics vary across persona types and
rate structures and (2) whether observed differences were statistically
significant. Descriptive analyses summarized the distribution of recommended
appliances and intervention strategies for each experimental condition. We
calculated the means and standard deviations for strategy type percentages,
targeted appliances, and persona fidelity scores, stratified by persona type,
utility rate structure, and their combinations. Inferential analyses employed
Chi-square tests of independence \cite{r70} to examine whether targeted
appliance patterns differed significantly across personas and rate types, with
Cram\'{e}r's V as the effect size measure \cite{r71}. We also employed one-way
analysis of variance (ANOVA) \cite{r72} to test the main effect of utility
rate structure on strategy type distribution and repeated measures ANOVA
\cite{r73} to test the main effect of persona type on strategy type
distribution, with household as the subject factor. This within-subjects
design controlled for between-household variability in energy use patterns and
appliance configurations, providing a more sensitive test of persona effects.

% ============================================================
\section{Results}
\label{sec:results}

\subsection{Empirical Validation Analysis}
\label{sec:emp-results}

Table~\ref{tab:performance} organizes the results of performance metrics,
derived from the three households. Overall, the proposed framework generated
interventions that aligned closely with reference interventions and utilized
the provided energy analysis data frequently and accurately.

\begin{figure}[!t]
  \centering
  \includegraphics[width=0.70\textwidth]{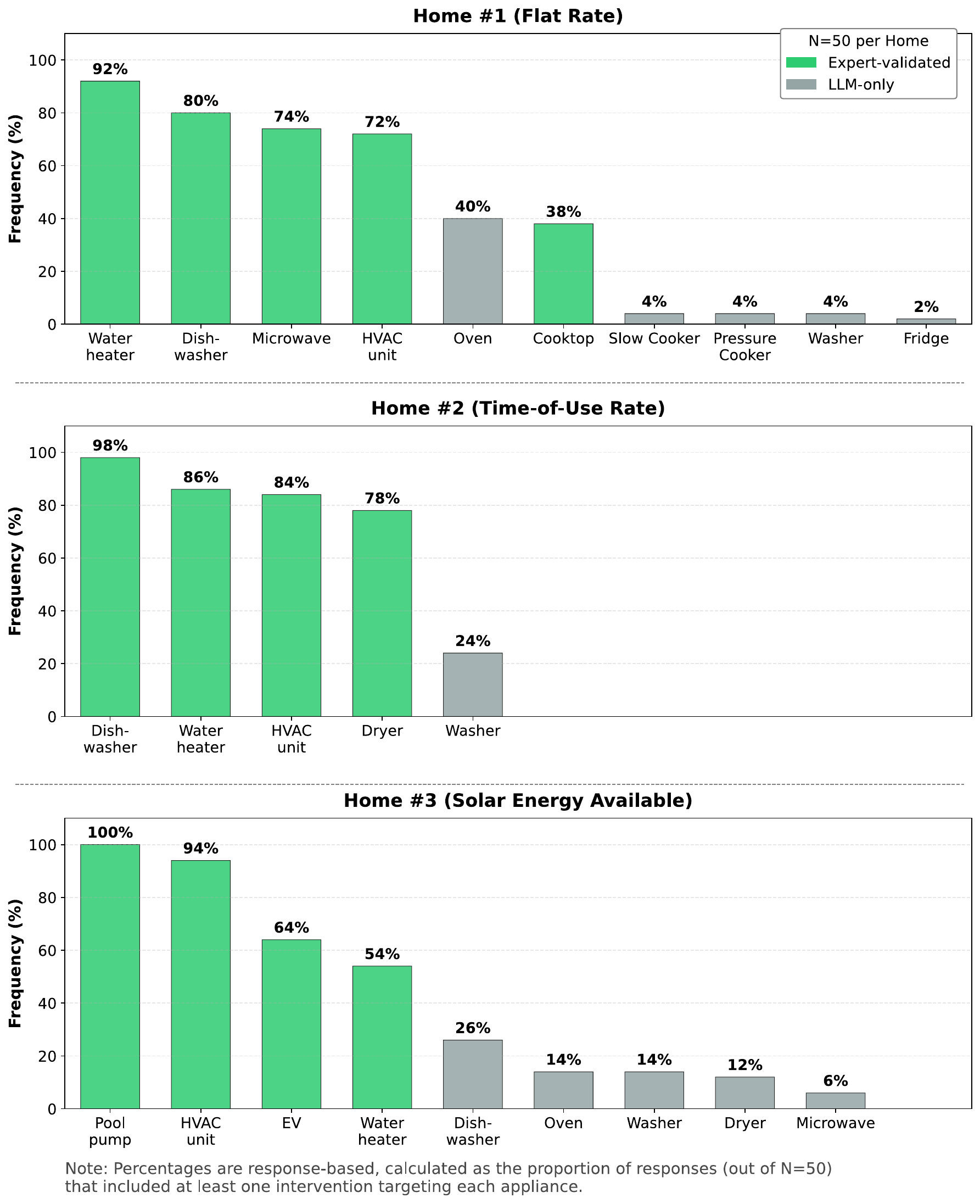}
  \caption{Appliances recommended through the proposed framework
  (response-level quantification)}
  \label{fig:appfreq}
\end{figure}

Reference alignment rates of 89.8--95.9\% (92.0\%, overall) demonstrated the
framework's ability to identify effective energy-saving strategies consistent
with reference solutions. Appliance validation rates ranged from 94.3\% to 100\%
(96.6\%, overall), indicating that the vast majority of generated interventions
targeted appliances actually present in each household's data rather than
hallucinated appliances (Figure~\ref{fig:appfreq}).

In Home~\#1 (flat-rate, no solar), the water heater was the most frequently
targeted appliance (92\% of responses), reflecting its position as the
second-largest energy consumer (4.2\% of total consumption,
Table~\ref{tab:A3}) combined with the curtailment-dominant strategy under
flat-rate pricing. The dishwasher (80\%), microwave (74\%), HVAC
unit (72\%), and cooktop (38\%) followed as secondary targets. The oven (40\%),
slow cooker, pressure cooker, washer, and fridge (each 2--4\%) represent
LLM-only recommendations for appliances not present in the household data, most
likely inferred from co-occurring appliance associations in training data
(for example, ovens and pressure cookers from the presence of a cooktop).

\begin{figure}[!t]
  \centering
  \includegraphics[width=0.65\textwidth]{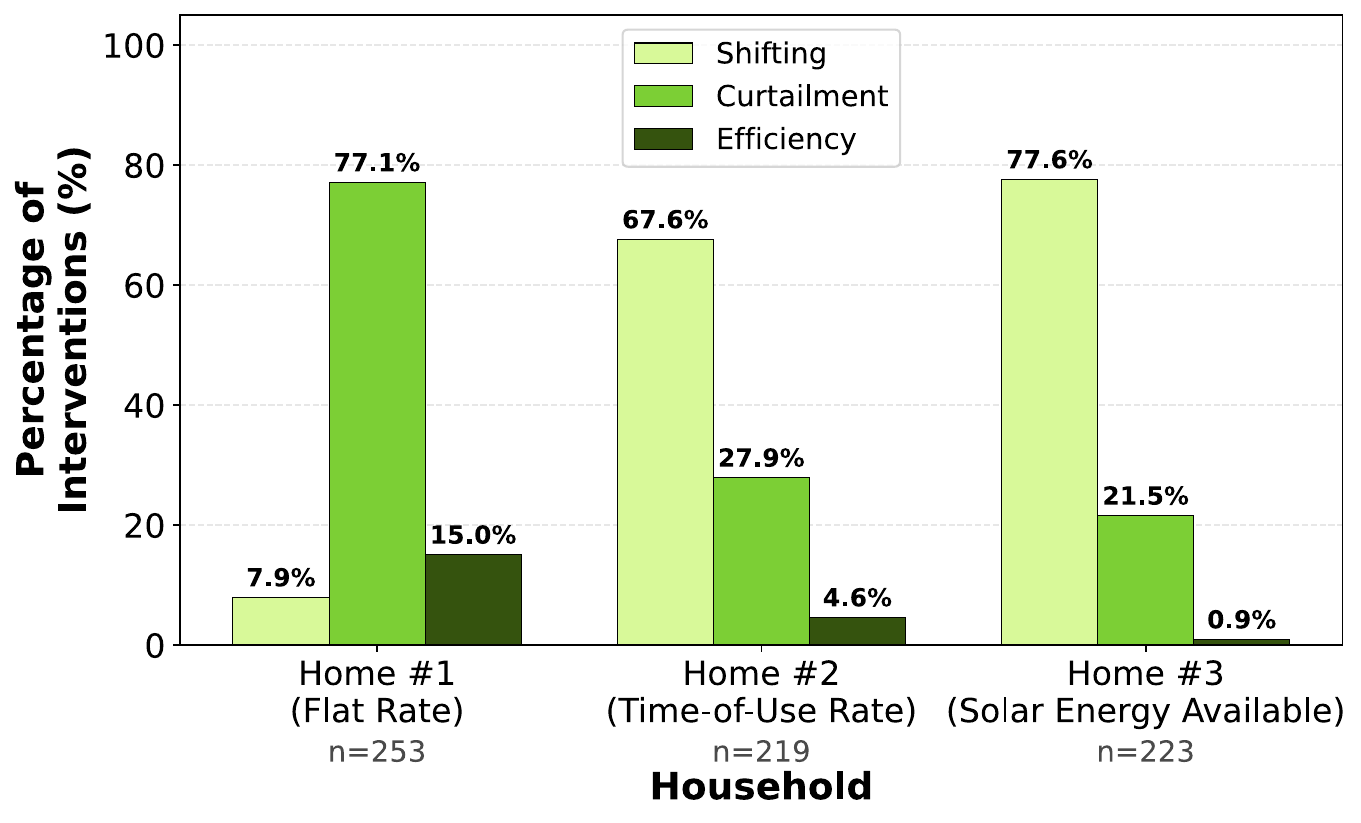}
  \caption{Intervention strategy distribution, recommended in three targeted
  households.}
  \label{fig:stratdist}
\end{figure}

In Home~\#2 (TOU-rate, environmentally motivated, no solar), the dishwasher
emerged as the dominant target (98\% of responses). Under TOU pricing, the dishwasher's high shiftability (it can be readily
scheduled to off-peak hours without lifestyle impact) made it the primary
vehicle for cost reduction, despite accounting for only 3.2\% of the
household's total consumption (Table~\ref{tab:A4}). The
water heater (86\%), HVAC unit (84\%), and dryer (78\%) followed, consistent
with their high energy contributions and shiftable load profiles. The
framework's strong preference for load-shifting strategies in this home (67.6\%,
Figure~\ref{fig:stratdist}) reflects appropriate recognition that temporal
redistribution of energy use yields greater financial benefit than simple
curtailment under TOU pricing.

In Home~\#3 (flat-rate with solar, dinner-routine constraint), the pool pump
was targeted in all 50 responses (100\%). This reflects the intersection of two
factors captured in the framework's solar alignment metrics: the pool pump is
Home~\#3's second-largest energy consumer (31.6\% of total consumption,
Table~\ref{tab:A5}) and exhibited a solar power alignment schedule of 0.60 and coverage
of 0.26, indicating significant potential to shift operation into periods of
high solar generation. The HVAC unit (94\%) and EV charger (64\%) followed
similarly, as both are high-power, schedulable loads with meaningful solar
alignment scores. The framework successfully identified solar self-consumption
as the primary optimization opportunity in this home, as evidenced by the
dominance of load-shifting strategies (77.6\%, Figure~\ref{fig:stratdist}).

\begin{table}[!t]
\centering
\caption{Performance evaluation results from the empirical validation
analysis}
\label{tab:performance}
\small
\begin{tabular}{llllll}
\toprule
& & & Home \#1 & Home \#2 & Home \#3 \\
\midrule
\multicolumn{3}{l}{Total number of interventions} & 248 & 229 & 265 \\
\midrule
\multirow{2}{*}{Accuracy}
  & \multicolumn{2}{l}{Appliance validity rate}  & 94.3\% & 100\%  & 96.0\% \\
  & \multicolumn{2}{l}{Reference alignment rate} & 90.5\% & 95.9\% & 89.8\% \\
\midrule
\multirow{10}{*}{Data citation}
  & \multirow{2}{*}{Appliance energy use}
    & Frequency & 96.5\% & 98.7\% & 92.7\% \\
  & & Accuracy  & 100\%  & 100\%  & 99.6\% \\
  & \multirow{2}{*}{Appliance usage patterns}
    & Frequency & 92.3\% & 98.7\% & 90.8\% \\
  & & Accuracy  & 90.8\% & 90.2\% & 97.2\% \\
  & \multirow{2}{*}{Utility rate}
    & Frequency & 18.8\% & 84.5\% & 49.6\% \\
  & & Accuracy  & 97.9\% & 100\%  & 91.7\% \\
  & \multirow{2}{*}{Household context}
    & Frequency & 68.9\% & 63.0\% & 14.7\% \\
  & & Accuracy  & 96.2\% & 97.0\% & 100\%  \\
  & \multirow{2}{*}{Solar analysis}
    & Frequency & N/A & N/A & 75.6\% \\
  & & Accuracy  & N/A & N/A & 87.3\% \\
\midrule
\multirow{3}{*}{Token efficiency}
  & \multicolumn{2}{l}{Input tokens} & 2,079 & 2,274 & 2,687 \\
  & \multicolumn{2}{l}{Mean output tokens} & 1,226 & 1,473 & 1,471 \\
  & \multicolumn{2}{l}{Average tokens per intervention} & 283.3 & 334.9 & 285.3 \\
\bottomrule
\end{tabular}
\end{table}

As illustrated in Figure~\ref{fig:stratdist}, the framework shifted its main
intervention strategies based on each household's utility rate structure,
available energy sources (i.e., solar energy), and their contextual
characteristics. For Home~\#1, curtailment dominated (77.1\%) since there was
no financial incentive to shift energy consumption to different times of day
--- reducing overall energy usage was the primary path to savings. Also, rather
than suggesting aggressive thermostat adjustments, responses noted that ``HVAC
efficiency must be well balanced with comfort''. In contrast, Homes~\#2 and
\#3, which had time-varying economic signals (TOU pricing and solar
availability, respectively), showed a strong preference for load shifting
strategies (67.6\% and 77.6\%). This indicates that the LLM properly recognized
when temporal redistribution of energy use would yield greater benefits than
simple reduction. Also, LLM-generated responses for Home~\#2, for example,
recognized the households' environmental priority, stating that ``the household
values sustainability and is willing to modify routines for energy savings.''
For Home~\#3, responses stated that ``they must ensure interventions do not
disrupt their dinner routine.'' The framework's ability to adapt its strategic
approach based on contextual factors, rather than applying a one-size-fits-all
recommendation pattern, demonstrates effective grounding in household-specific
data.

A concrete example further clarifies the nature of this contextual
adaptation. Home~\#3 carries both a solar generation context and a 6~PM
weekday dinner routine, allowing simultaneous multi-constraint reasoning to
be observed directly. The pool pump (31.6\% of total consumption, solar alignment
score 0.60) was recommended for early morning (6--8~AM) or late evening
(7--9~PM) as the primary scheduling window, with daytime solar hours
(10~AM--3~PM) as an alternative when feasible, a schedule that avoids the
dinner-time window while preserving solar self-consumption opportunity. The EV
charger (4.1\%), which carries no dinner-time conflict, received an
unconditional recommendation to charge during peak solar hours (10~AM--3~PM).
The framework thus applied two distinct scheduling rationales to two appliances
within the same household, reflecting the simultaneous influence of solar
availability and routine constraint without requiring explicit per-appliance
rules.

Data citation patterns revealed both strengths and notable variations across
data categories. Appliance energy use and usage patterns were cited most
frequently (90.8--98.7\%) with high accuracy (90.2--100\%), indicating strong
grounding in the energy consumption data. Utility rate citation frequency
varied substantially across households, from 18.8\% in Home~\#1 (flat rate)
to 84.5\% in Home~\#2 (time-of-use rate), reflecting appropriate adaptation
to rate structure relevance. When utility rates were cited, accuracy
remained high (91.7--100\%). Household context citation frequency showed
considerable variation (14.7--68.9\%), suggesting inconsistent incorporation of
lifestyle and demographic factors. Solar analysis data, available only for
Home~\#3, was cited in 75.6\% of interventions with 87.3\% accuracy,
demonstrating the model's capability to integrate renewable energy
considerations when relevant data is provided.

Input token counts ranged from 2,079 to 2,687, reflecting the varying
characteristics of household data; Home~\#3 required the most context as its
14 appliances including solar generation. Output tokens remained relatively
consistent (1,226--1,473), suggesting stable response generation patterns.
Average tokens per intervention (283--335) indicates that each recommendation
included substantial reasoning and justification beyond the core feasible
advice, as directed by the SC-CoT prompt. The slightly higher tokens per
intervention in Home~\#2 reflected more detailed TOU scheduling explanations
required for its hierarchical rate structure.

\subsection{Combinatorial Analysis}
\label{sec:comb-results}

\begin{figure}[!t]
  \centering
  \includegraphics[width=0.80\textwidth]{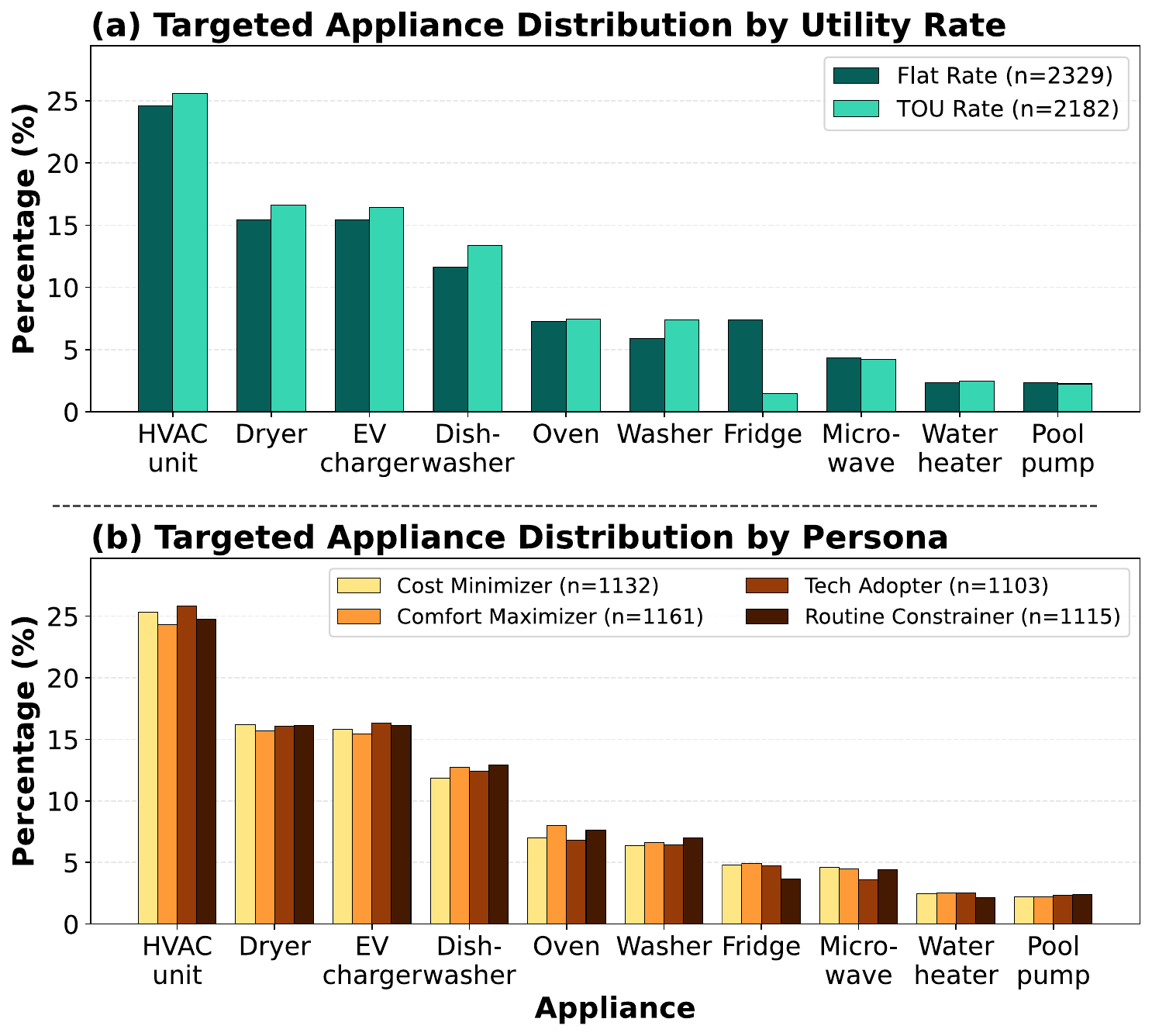}
  \caption{Targeted appliance distribution by (a) rate type and (b) persona}
  \label{fig:appliancedist}
\end{figure}

Figure~\ref{fig:appliancedist} presents the distribution of appliances targeted
by the LLM-generated interventions, stratified by utility rates and personas.
Subfigure (a) revealed a noticeable divergence in refrigerator targeting between
rate structures and it led to a significant difference (Table~\ref{tab:chisq}).
Specifically, with TOU rates, it was much less targeted by the framework (flat
rates: 172 times and TOU rates: 32 times) and the refrigerator's chi-square
statistic was 87.1, accounting for 85.3\% of the total effect. Washing machines
and dishwashers were more frequently targeted under TOU rate structures due to
their shiftability.

\begin{table}[!htbp]
\centering
\caption{Chi-square test results for targeted appliance distribution
by (1) utility rate and (2) persona}
\label{tab:chisq}
\small
\begin{tabular}{lllll}
\toprule
Factor & Test & Chi-square ($\chi^2$) & P-value & Cram\'{e}r's V \\
\midrule
Utility rate & Chi-square test of independence & 102.19 & $<$0.001$^{*}$ & 0.151 \\
Persona      & Chi-square test of independence & 21.11  & 0.9997         & 0.040 \\
\bottomrule
\multicolumn{5}{l}{\footnotesize $^{*}$Statistically significant}
\end{tabular}
\end{table}

In contrast, persona did not play a role in differing targeted appliances. This
persona factor influenced how recommendations were framed and emphasized (e.g.,
prioritizing cost-savings for cost-minimizers), but did not alter which
appliances were optimal to recommend.

\begin{figure}[!t]
  \centering
  % Figure11_strategy_distribution_combined.svg is Figure 11
  \includegraphics[width=0.65\textwidth]{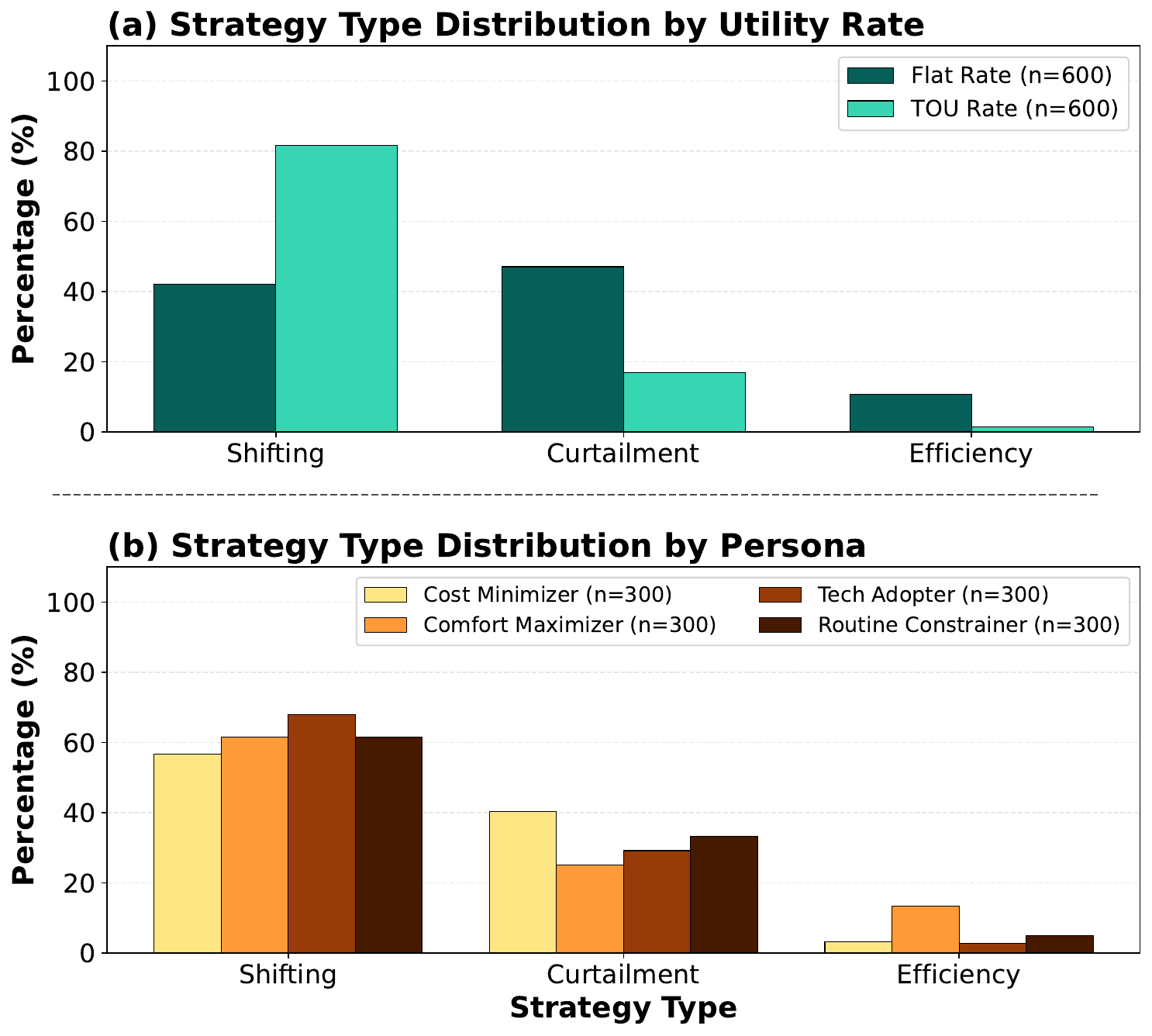}
  \caption{Strategy type distribution, stratified by (a) utility rate and
  (b) persona}
  \label{fig:strattype}
\end{figure}

Figure~\ref{fig:strattype} illustrates the distribution of energy-saving
strategy types generated by the proposed framework across two utility rate
structures and personas. Subfigure~(a) demonstrates a significant rate structure
effect on strategy distribution (Table~\ref{tab:anova}). Under flat rate
pricing, interventions were relatively balanced between shifting and curtailment,
with minimal efficiency recommendations. The reason for a substantial proportion
of load-shifting interventions, even with flat rate structures, was that 58\%
of households owned solar panels. In contrast, TOU rate pricing drove a strong
preference for shifting strategies, while curtailment and efficiency
recommendations decrease substantially. Subfigure~(b) reveals persona-based
variation in strategy distributions. All four personas revealed majority
shifting recommendations, consistent with the strong influence of utility rate
structure. However, notable differences emerged in secondary strategies: Cost
minimizers showed higher curtailment preference, while comfort maximizers
demonstrated elevated curtailment and efficiency recommendations compared with
cost minimizers' efficiency rate. Tech adopter and routine constrainer exhibited
similar patterns, with minimal efficiency recommendations across all personas.
Such variations resulted in statistically significant differences
(Table~\ref{tab:anova}).

A concrete case illustrates how these aggregate differences manifest in
individual recommendations. Holding the household constant---the household
designated Home~\#1 in the empirical validation analysis, whose dominant
appliance is HVAC at 92.3\% of total energy use---and varying only the
persona demonstrates that persona conditioning changes the substance of
recommendations rather than only their framing. Under the cost-minimizer
persona, the framework recommended adjusting the thermostat setpoint across
a wide peak-hour window (10~AM--10~PM), with expected savings of
307.7~kWh. Under the routine-constrainer persona, the same adjustment was
temporally restricted with an explicit compatibility check:
\textit{``Ensure temperature changes occur during sleeping hours or when
the house is less occupied.''} A parallel contrast appeared in the
dishwasher recommendation: the cost-minimizer received a savings-focused
directive (full loads, air drying, 6.1~kWh reduction) without timing
specification, while the routine-constrainer received an otherwise
identical directive anchored explicitly to schedule: \textit{``Run after
dinner, ensuring it aligns with family routines.''}

\begin{table}[!htbp]
\centering
\caption{ANOVA test results for strategy type distribution by utility
rate and persona}
\label{tab:anova}
\small
\begin{tabular}{llllll}
\toprule
Factor & Test & F-statistic & P-value & Effect size ($\eta^2$) \\
\midrule
Utility rate & One-way ANOVA & 651.65 & $<$0.0001$^{*}$ & 0.35 \\
Persona & Repeated measures ANOVA & 11.43 & $<$0.0001$^{*}$ & 0.09 \\
\bottomrule
\multicolumn{5}{l}{\footnotesize $^{*}$Statistically significant}
\end{tabular}
\end{table}

\begin{figure}[!t]
  \centering
  \includegraphics[width=0.60\textwidth]{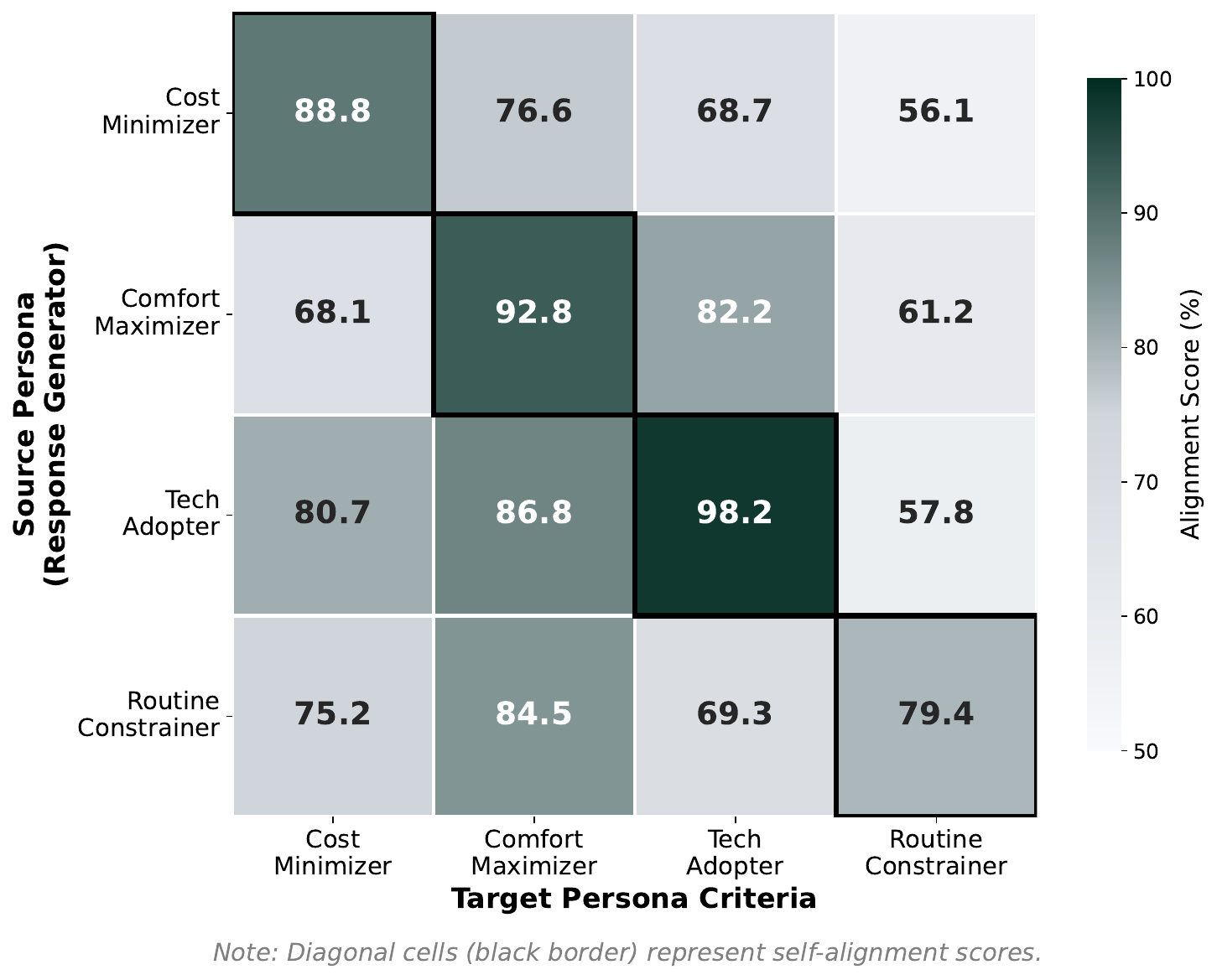}
  \caption{Cross-persona fidelity matrix}
  \label{fig:fidelity}
\end{figure}

The cross-persona fidelity matrix (Figure~\ref{fig:fidelity}) reveals that
LLM-generated responses exhibited distinct behavioral signatures aligned with
their assigned personas. All four personas achieved high self-alignment scores
on the diagonal: Tech adopter (98.2\%), comfort maximizer (92.8\%), cost
minimizer (88.8\%), and routine constrainer (79.4\%). The mean diagonal score
of 89.8\% against a mean off-diagonal score of 72.3\% yields an overall
discrimination gap of 17.5 percentage points, confirming that the framework
produced meaningfully differentiated behavioral profiles across the four
personas.

Each cell in the matrix reports the alignment score when responses generated 
under the \textit{source} persona (row) are evaluated against the criteria 
defined for the \textit{target} persona (column). Diagonal cells (outlined 
in black) represent self-alignment: how faithfully each persona's responses 
satisfy its own definitional criteria. Off-diagonal cells capture the degree 
to which a persona's recommendations inadvertently satisfy the criteria of 
a different persona, reflecting behavioral proximity or conceptual overlap 
between persona types.

Among diagonal scores, the Tech-adopter achieved the highest self-alignment
(98.2\%), indicating that its distinctive emphasis on automation, scheduling,
and smart-device language was strongly and consistently elicited by the
corresponding prompt conditioning. The Routine-constrainer achieved the lowest
diagonal score (79.4\%), attributable to the inherently implicit nature of
routine preservation: recommendations that avoid disruption do not always name
specific constraints explicitly, making the assessment of ``schedule inquiry
proactivity'' (Table~\ref{tab:criteria}, Criterion~3) more variable across runs.

Several off-diagonal patterns reveal behavioral proximities worth noting. The
Routine-constrainer scored 84.5\% against the Comfort-maximizer criteria, the
highest off-diagonal score in the matrix. This is expected: both personas share
a preference for non-intrusive interventions, and recommendations that avoid
disrupting routines naturally tend to also avoid comfort sacrifices. Conversely,
the Cost-minimizer scored only 56.1\% against the Routine-constrainer criteria,
the lowest off-diagonal value, reflecting a meaningful behavioral divergence:
cost-focused recommendations that accept comfort tradeoffs in pursuit of savings
are fundamentally misaligned with routine-preserving criteria that prioritize
schedule fidelity. The Comfort-maximizer scored 82.2\% against the Tech-adopter
criteria: comfort-preserving recommendations frequently favor efficiency
upgrades (e.g., smart thermostats, programmable setpoints) over curtailment,
producing incidental alignment with automation-oriented criteria. The reverse
relationship was weaker (68.1\%), since technology-driven recommendations do not
inherently address thermal comfort preferences.

% ============================================================
\section{Discussion and Limitation}
\label{sec:discussion}

The findings of this study suggest that the proposed framework can offer a
promising pathway toward more adaptive, effective, and feasible household energy
guidance. A key element of this framework was the LLM, whose strength lies in
its ability to accurately comprehend the semantic meaning of structured prompts
that encode heterogeneous household energy analysis results and contextual
information. Rather than relying solely on statistical patterns, the LLM enabled
higher-level reasoning by translating quantitative energy metrics and qualitative
contextual information into coherent, human-readable eco-feedback that was both
context-aware and feasible, highlighting the LLM's role as a reasoning and
synthesis engine rather than a simple text generator \cite{r78}.

Another key element was the energy analysis component (illustrated in
Figures~\ref{fig:framework} and~\ref{fig:validation}), which played a
substantial role in extracting quantitative insights from household energy data.
By externalizing core analytical tasks, this component ensured analytical rigor,
consistency, and interpretability, while reducing the computational and reasoning
burden allocated on the LLM. This system architecture aligned with the findings
from our previous study \cite{r59}, which demonstrated the benefits of
separating deterministic energy analysis from language-based reasoning. In
contrast, when raw energy data were directly provided to the LLM without prior
analytical structuring, the performance of such frameworks became constrained by
token limitations (which restricted the volume of data that could be processed)
and by the inherent stochastic nature of LLMs, which does not guarantee
consistent or reproducible analytical procedures across repeated evaluations.

The choice to assign intervention synthesis, rather than communication,
to the LLM reflects the nature of the integration task. A rule-based approach
would require predefined decision rules for each meaningful combination of
appliance profiles, utility rate conditions, solar availability, and
persona-defined behavioral constraints. Across the 50-household dataset with
two rate structures and four personas, this combinatorial space is large, and
fixed rules cannot naturally accommodate the contextual nuance that
distinguishes, for example, a cost-minimizer's willingness to accept moderate
comfort tradeoffs from a routine-constrainer's strict avoidance of any change
to dinner-time appliance use. The LLM's capacity to reason over structured
textual inputs and produce coherent, contextually differentiated
natural-language guidance is particularly suited to this multi-factor synthesis
task. The significant persona effect observed in strategy distribution
($F = 11.43$, $p < 0.0001$, $\eta^2 = 0.09$) and the differentiated
cross-persona fidelity matrix (Figure~\ref{fig:fidelity}) provide empirical
evidence that the framework produced meaningfully distinct recommendations
across behavioral profiles, consistent with LLM-based contextual synthesis
rather than generic template filling. The combinatorial nature of the
integration task (household-specific appliance profiles, utility rate
incentives, solar availability, and persona-defined behavioral constraints
interacting simultaneously) makes a rule-based encoding of equivalent
contextual nuance intractable, further motivating the LLM's synthesis role.

While this study processes accumulated historical appliance-level data to generate
periodic eco-feedback, the contextual engineering approach underlying the framework
is not constrained to retrospective operation. The core architecture
(deterministic energy analysis feeding a structured LLM prompt) transfers
directly to real-time pipelines where rolling or streaming consumption data
replace accumulated historical records. This transferability has been
demonstrated in our work: \cite{r77} extends this approach to a fully
conversational, real-time multi-agent home energy management assistant (HEMA),
in which the same separation of deterministic analysis and LLM-based synthesis
supports live energy queries, context-aware recommendations, and appliance
scheduling across a multi-agent system. The retrospective mode adopted in this
study reflects the requirements of the batch validation design (enabling
systematic performance evaluation across all proposed metrics over a complete
dataset), rather than a structural constraint of the framework.

A potential key player is building information. Building-related attributes such
as floor area, construction vintage, envelope characteristics, insulation levels,
window-to-wall ratio, HVAC system type, and overall thermal performance can
substantially influence household energy consumption patterns and the feasibility
of different energy-saving interventions. Incorporating such information could
further enhance the contextual grounding of eco-feedback by enabling the
framework, for example, to distinguish between energy use driven by structural
constraints versus occupant behavior, hence supporting more targeted and
realistic recommendations.

A methodological limitation of this study was the use of synthesized household
characteristics. They might not fully capture the complexity and diversity of
real households. To mitigate this limitation, we constructed household
characteristics by grounding them in appliance usage patterns and literature,
aiming to approximate realistic household contexts as closely as possible. Also,
the household energy data were drawn from the summer season of Austin, Texas
(June to August), which might limit the generalizability of the interventions
applicable to other seasons or climate conditions, particularly those dominated
by heating-driven energy use.

% ============================================================
\section{Conclusion}
\label{sec:conclusion}

This study examined the potential of generating context-aware eco-feedback
through the proposed context engineering-based framework and validated their
accuracy, data utilization, and adaptivity using systematic empirical validation
and combinatorial analyses. The results showed consistently strong performance
across all evaluation metrics: the LLM-generated eco-feedback accurately
identified high-impact appliances and proposed contextually reasonable
energy-saving interventions while effectively reflecting household-specific
characteristics. The framework incorporated systematic and quantitative
evaluations of appliance-level energy use data from households and qualitative
assessments of contextual factors, including personas and non-negotiable
routines. Due to this integrated analytical lens, the generated eco-feedback
were more adaptive, effective, and feasible across a wide range of household
contexts. Collectively, these findings advance the development of context-aware
interactions between occupants and buildings, laying a foundation for improved
occupant well-being and more sustainable building energy management.

The empirical validation demonstrated that the framework reliably interpreted
household-specific contextual data and generated eco-feedback consistent with
expert-derived reference solutions. Reference alignment rates of 89.8--95.9\%
(92.0\% overall) and appliance validation rates of 94.3--100\% (96.6\% overall)
confirmed that the generated interventions targeted real household appliances
and matched established energy-saving strategies. Data citation accuracy of
95.7\% indicated that the model drew consistently and correctly on the provided
household data rather than producing generic advice. The combinatorial analysis
further demonstrated the framework's adaptability: the persona fidelity
self-alignment scores ranged from 79.4\% to 98.2\% (mean diagonal: 89.8\%),
against a mean off-diagonal score of 72.3\%, yielding a 17.5-percentage-point
discrimination gap that confirms the framework produced meaningfully
differentiated behavioral recommendations across household contexts. Utility
rate structure exerted a large main effect on intervention strategy distribution
($F = 651.65$, $p < 0.0001$, $\eta^2 = 0.35$), while persona type produced a
statistically significant medium effect ($F = 11.43$, $p < 0.0001$,
$\eta^2 = 0.09$), collectively demonstrating that the framework responded
appropriately to both the economic and behavioral dimensions of household
context.

As part of our future research, we plan to systematically assess the
effectiveness of different context-aware feedback delivery types through controlled experimental studies and field studies,
examining how variations in feedback modality,
timing, and presentation influence household engagement, comprehension, and
sustained energy-saving behaviors. The proposed framework was focused more on \textit{generating}
context-aware eco-feedback, rather than \textit{interacting} with users,
which is the key aspect in the eco-feedback eco-system.
% ============================================================
\credit{\textbf{Wooyoung Jung}: Writing -- review \& editing, Writing -- original
draft, Visualization, Validation, Supervision, Software, Resources, Project
administration, Methodology, Investigation, Funding acquisition, Formal
analysis, Data curation, Conceptualization; \textbf{Prosper Babon-Ayeng}:
Methodology, Investigation, Formal analysis, Data curation.}

\acknowledgments{This material is based upon work partially supported by Salt River Project
(SRP) under grant SRP UA-08 24-25 and SRP UA-11 25-26 and the National Science
Foundation (NSF) under grant \#2519054. Any opinions, findings, and conclusions
or recommendations expressed in this material are those of the authors and do
not necessarily reflect the views of the NSF or SRP. We would like to
appreciate our SRP advisor, Victor J.\ Berrios, who provided external evaluation
of our methodological approach.}

\aiuse{During the preparation of this work the authors employed ChatGPT 5, 5.1, and
5.2, and Claude Sonnet 4, 4.5 in order to proofread sentences and write Python
codes for analyses and visualization. After using these tools/services, the
authors reviewed and edited the content as needed and take full responsibility
for the content of the publication.}

\competing{The authors declare that they have no known competing financial interests or
personal relationships that could have appeared to influence the work reported
in this paper.}

% ============================================================
% APPENDICES
% ============================================================
\appendix
\startappendix

\section*{Appendix}

\section{Household Appliance-Level Energy Use Data}
\label{appx:data}

This section explains the household appliance-level energy use data of 50
households. The dataset represents diverse energy consumption patterns,
technology adoption profiles, and appliance ownership configurations, capturing
a wide spectrum of residential energy behaviors.

\begin{itemize}
  \item \textbf{Technology adoption}: The technology distribution of 50
        households reflected increasing trends in renewable energy and EV
        adoption, with 58\% of households having solar panels and 56\% owning
        EV chargers. Table~\ref{tab:tech} presents the detailed breakdown of
        technology adoption across the dataset.
\end{itemize}

\begin{table}[!htbp]
\centering
\caption{Technology adoption distribution across the 50-household
dataset}
\label{tab:tech}
\small
\begin{tabular}{ll}
\toprule
Technology adoption category & Number of households (\%) \\
\midrule
With solar panels and EVs & 18 (36.0\%) \\
With solar panels         & 11 (22.0\%) \\
With EVs                  & 10 (20.0\%) \\
Without solar panels or EVs & 9 (18.0\%) \\
Battery storage system    & 2 (4.0\%) \\
\bottomrule
\end{tabular}
\end{table}

\begin{itemize}
  \item \textbf{Energy consumption profile}: As summarized in
        Table~\ref{tab:energystats}, total energy consumption in the dataset
        exhibited wide variability, ranging from 206.73~kWh to 8,959.14~kWh
        over the monitoring period (21 homes had three months of data and 29
        had one month of data). The mean energy consumption was 1,449.83~kWh
        (standard deviation: 1,700.73~kWh), while the median was 799.11~kWh.
        Notably, households with both solar PV and EVs showed the highest mean
        consumption (2,112.97~kWh), while households with EVs exhibited the
        lowest mean consumption (659.44~kWh).
\end{itemize}

\begin{table}[!htbp]
\centering
\caption{Energy consumption statistics by technology adoption
category}
\label{tab:energystats}
\small
\begin{tabular}{llll}
\toprule
Technology adoption category & Mean (kWh) & Median (kWh) & Range (kWh) \\
\midrule
With solar panels and EVs   & 2,113.0 & 855.1 & 367.6 -- 8,959.1 \\
With solar panels           & 1,413.1 & 1,534.4 & 419.8 -- 2,461.0 \\
Without solar panels or EVs & 1,211.6 & 836.0 & 206.7 -- 4,623.3 \\
With battery storage systems & 707.6  & 707.6 & 653.1 -- 762.1 \\
With EVs                    & 659.4   & 628.2 & 406.6 -- 1,003.1 \\
\bottomrule
\end{tabular}
\end{table}

\begin{itemize}
  \item \textbf{Appliance distribution and characteristics}: The number of
        monitored appliances per household varied considerably, ranging from 2
        to 21 appliances, with a mean of 9.3 appliances (SD = 5.1) and a
        median of 10 appliances.
\end{itemize}

\section{Household-Specific Reference Eco-Feedback}
\label{appx:reference}

This section elaborates on how the reference energy-saving interventions were
derived from the reasoning processes described in Section~\ref{sec:reasoning}.
Again, these strategies were household-specific interventions.

\textbf{Home \#1}: Since this household subscribed to the flat rate structure
and did not generate on-site solar power, energy saving opportunities were
limited to load curtailment strategies as follows:

\begin{itemize}
  \item \textbf{HVAC unit}: dominated household electricity consumption
        (Table~\ref{tab:A3}), clearly identifying it as the primary target for
        curtailment-based interventions in spite of the household persona. It
        exhibited high operating frequency during peak hours as well as elevated
        variability across both timeframes, reflecting the household's
        temperature sensitivity and continuous reliance on thermal conditioning.
        Accordingly, the potential energy-saving interventions include:
        (1) installing smart thermostats to automatically adjust temperature
        setpoints during vacancy; (2) modestly widening temperature setpoints
        while supplementing comfort with local systems (e.g., ceiling or
        portable electric fans); or (3) managing infiltration and exfiltration
        through improved door and window operation, sealing air leaks, or
        adopting weatherization practices.
  \item \textbf{Electric water heater}: represented a distinct secondary
        contributor to household energy usage and its curtailment strategies
        include: (1) lowering water temperature setpoints; (2) reducing shower
        duration or installing low-flow showerheads; (3) addressing hot-water
        leakage (if any) and improving pipe or tank insulation; or (4) using
        cold water for laundry when feasible.
  \item \textbf{Cooktop and Microwave}: emerged as the fourth and fifth largest
        contributors, indicating that meal preparation often occurred at home.
        Efficiency gains can be achieved using behavioral adjustments. For the
        cooktop, recommended strategies include matching burner size to cookware,
        using lids to shorten cooking times, or leveraging residual heat by
        turning burners off before food is fully cooked. For the microwave,
        energy savings can be realized by defrosting in refrigerator overnight
        instead of microwave or minimizing reheating cycles.
  \item \textbf{Dishwasher}: was used intermittently following meal preparation
        (and dining activities, possibly). Several operational strategies can
        improve efficiency. These include running the dishwasher only when fully
        loaded; reducing usage frequency from daily to every other day; skipping
        the heated dry cycle; thoroughly scraping dishes prior to loading reduce
        required wash intensity.
\end{itemize}

\begin{table}[!htbp]
\centering
\caption{Appliance-level energy use analysis for Home \#1}
\label{tab:A3}
\small
\begin{tabular}{lllll}
\toprule
Appliance & \multicolumn{2}{l}{Energy use} & Frequency & Variability \\
          & Total (kWh) & Mean (kW) & (daily) & (daily) \\
\midrule
HVAC unit             & 769.22 (92.3\%) & 1.42 & 0.36 & 0.66 \\
Electric water heater & 34.83 (4.2\%)   & 1.15 & 0.02 & 1.93 \\
Cooktop               & 9.45 (1.1\%)    & 0.45 & 0.01 & 1.71 \\
Microwave             & 8.57 (1.0\%)    & 0.40 & 0.01 & 1.08 \\
Dishwasher            & 7.59 (0.9\%)    & 0.48 & 0.01 & 1.74 \\
\bottomrule
\multicolumn{5}{l}{\footnotesize Note: This table excluded appliances whose
total energy consumption was less than 0.5\%}
\end{tabular}
\end{table}

\textbf{Home \#2}: This household subscribed to a TOU rate structure, which
introduced clear financial incentives for both load curtailment and load
shifting. Given the household's strong motivation toward energy savings, no
behavioral or comfort-related constraints were imposed when formulating
strategies. Analysis identified four appliances with meaningful energy
consumption contributions (Table~\ref{tab:A4}), for which targeted
energy-saving interventions were subsequently developed, as described below:

\begin{itemize}
  \item \textbf{HVAC unit}: was the dominant energy consumer and showed high
        concentration during the peak hours, making it the major target for
        both curtailment and load-shifting interventions. In addition to the
        strategies identified in Home~\#1, more aggressive measures can be
        considered, e.g., actively widening temperature setpoints up to 3--4\textdegree{}C.
        Additional strategies include pre-cooling the home two to three hours
        before peak hours to reduce peak demand and broadening the thermostat
        deadband to limit compressor cycling during peak hours.
  \item \textbf{Electric water heater}: emerged as the second largest
        contributor and showed consistent running throughout the day. Building
        on the strategies identified in Home~\#1, additional load-shifting
        interventions could be considered to concentrate its usage during
        off-peak hours, e.g., scheduling water heating primarily during
        off-peak hours or maintaining only minimal heating during peak periods
        to ensure service continuity.
  \item \textbf{Electricity-powered clothes dryer}: This clothes dryer appeared
        to be the third energy consumer in the household, but with highest mean
        power. It was more often utilized during peak hours; hence, possible
        interventions include air-drying clothes using outdoor lines or indoor
        drying racks, limiting dryer use to essential loads only; or scheduling
        dryer operation exclusively during off-peak hours.
  \item \textbf{Dishwasher}: was often employed during peak hours and recorded
        as the fourth energy consumer. In addition to load curtailment
        interventions stated in Home~\#1, a load-shifting strategy (e.g.,
        running it exclusively during off-peak hours) can be considered.
\end{itemize}

\begin{table}[!htbp]
\centering
\caption{Appliance-level energy use analysis for Home \#2}
\label{tab:A4}
\small
\begin{tabular}{lp{1.8cm}p{1.5cm}p{3.2cm}p{3.2cm}}
\toprule
Appliance & Energy use Total (kWh) & Mean (kW) &
  Frequency (daily, off-, on-peak) &
  Variability (daily, off-, on-peak) \\
\midrule
HVAC unit             & 706.75 (84.8\%) & 1.84 & 0.25, 0.10, 0.71 & 1.17, 1.44, 0.17 \\
Electric water heater & 60.97 (7.3\%)   & 0.46 & 0.09, 0.09, 0.09 & 0.55, 0.62, 0.30 \\
Electricity-powered clothes dryer & 31.66 (3.8\%) & 2.08 & 0.01, 0.01, 0.02 & 1.20, 1.70, 0.32 \\
Dishwasher            & 26.97 (3.2\%)   & 0.67 & 0.03, 0.02, 0.06 & 0.95, 1.12, 0.29 \\
\bottomrule
\multicolumn{5}{l}{\footnotesize Note: This table excluded appliances whose
total energy consumption was less than 0.5\%}
\end{tabular}
\end{table}

\textbf{Home \#3}: This household combined a flat rate structure with solar
power generation, creating optimization opportunities focused on maximizing
solar self-consumption. The load shifting interventions should target the
hours when solar energy is available. Analysis identified five appliances
with meaningful energy consumption contributions (Table~\ref{tab:A5}), of
which three were selected for targeted interventions based on their
combination of consumption share, usage variability, and solar alignment
potential, as described below:

\begin{itemize}
  \item \textbf{HVAC unit}: was again the major energy contributor for this
        household. Similar load curtailment and shifting strategies identified
        in Homes~\#1 and \#2 remained applicable, but shifting interventions
        were explicitly constrained to periods of on-site solar availability.
  \item \textbf{Pool pump}: emerged as the second dominant energy consumer,
        presenting clear opportunities for both curtailment and load shifting,
        especially given high variability. Curtailment solutions include
        reducing overall runtime, cleaning filters regularly to lower pumping
        load, optimizing pool chemistry to reduce filtration demand, and
        covering the pool when not in use to minimize debris accumulation.
        Load-shifting strategies focus on scheduling pump operation during
        periods of high solar generation, through manual scheduling or
        automated timers when available.
  \item \textbf{EV charger}: appeared as the third energy contributor,
        characterized by high mean power, low frequency, and high variability.
        This operating profile indicates great potential for targeted load
        curtailment and shifting strategies. Charging events can be curtailed
        by optimizing driving efficiency, activating eco-driving mode, or
        combining errands to reduce driving frequency. Additionally, charging
        can be shifted to periods of high on-site solar generation or off-peak
        grid hours, leveraging smart charging controls or manual scheduling to
        minimize grid reliance and reduce electricity costs without compromising
        mobility needs.
  \item \textbf{Electric water heater}: emerged as the fourth-largest energy
        consumer in this household. Energy-saving strategies identified in
        Homes~\#1 and \#2 remained applicable; however, they should target
        periods of on-site solar availability to enhance self-consumption while
        maintaining essential hot-water service.
\end{itemize}

\begin{table}[!htbp]
\centering
\caption{Appliance-level energy use analysis for Home \#3}
\label{tab:A5}
\small
\begin{tabular}{lp{1.8cm}p{1.1cm}p{1.5cm}p{1.5cm}p{1.5cm}p{1.5cm}}
\toprule
Appliance & Energy use Total (kWh) & Mean (kW) & Frequency (daily) &
  Variability (daily) & Solar Power Alignment Schedule & Solar Power Alignment Coverage \\
\midrule
HVAC unit             & 5,248.97 (58.0\%) & 3.48 & 1.00 & 0.004 & 0.54 & 0.24 \\
Pool pump             & 2,858.41 (31.6\%) & 2.19 & 0.86 & 0.32  & 0.60 & 0.26 \\
EV charger            & 371.93 (4.1\%)    & 4.91 & 0.05 & 0.74  & 0.42 & 0.10 \\
Electric water heater & 366.48 (4.1\%)    & 1.16 & 0.21 & 0.49  & 0.40 & 0.14 \\
Refrigerator          & 120.65 (1.3\%)    & 0.08 & 1.00 & 0.00  & 0.52 & 0.23 \\
\bottomrule
\multicolumn{7}{l}{\footnotesize Note: This table excluded appliances whose
total energy consumption was less than 0.5\%}
\end{tabular}
\end{table}

\end{document}